\documentclass[12pt,preprint]{aastex} 
\shorttitle{Hot Subdwarf Companion to the Be Star FY~CMa} 
\shortauthors{Peters et al.} 
 
\begin{document} 
 
\received{2008 March 21} 
\accepted{} 
 
\title{Detection of a Hot Subdwarf Companion to the Be Star FY~Canis Majoris}  
 
\author{Geraldine J. Peters\altaffilmark{1,2}} 
\affil{Space Sciences Center, University of Southern California, Los Angeles, CA 90089-1341;  
gjpeters@mucen.usc.edu} 
 
\author{Douglas R. Gies\altaffilmark{1}, Erika D. Grundstrom\altaffilmark{1,3}} 
\affil{Center for High Angular Resolution Astronomy and  
 Department of Physics and Astronomy,\\ 
 Georgia State University, P. O. Box 4106, Atlanta, GA 30302-4106;  
 gies@chara.gsu.edu, erika.grundstrom@vanderbilt.edu} 
 
\author{M. Virginia McSwain\altaffilmark{1}} 
\affil{Department of Physics, Lehigh University, Bethlehem, PA 18105; 
 mcswain@lehigh.edu
} 
 
\altaffiltext{1}{Visiting Astronomer, Kitt Peak National Observatory, 
National Optical Astronomy Observatory, operated by the Association 
of Universities for Research in Astronomy, Inc., under contract with 
the National Science Foundation.} 
\altaffiltext{2}{Guest Observer with the International Ultraviolet Explorer
Satellite.} 
\altaffiltext{3}{Current address: Department of Physics and Astronomy, 
Vanderbilt University, Nashville, TN 37206.}

 
 
\begin{abstract} 
The rapid rotation of Be stars may be caused in some cases 
by past mass and angular momentum accretion in an interacting 
binary in which the mass donor is currently viewed as a small, hot
subdwarf stripped of its outer envelope. Here we report on the 
spectroscopic detection of such a subdwarf in the Be binary system 
FY~Canis Majoris from the analysis of data acquired by the {\it IUE} 
spacecraft and KPNO Coud\'{e} Feed Telescope over the course of 
16 and 21 years, respectively. 
We present a double-lined spectroscopic orbit for the binary
based upon radial velocities from the {\it IUE} spectra 
and use the orbital solutions with a Doppler tomography 
algorithm to reconstruct the components' UV spectra. 
The subdwarf is hot ($T_{\rm eff} = 45\pm5$~kK) and has a mass of
about $1.3 M_\odot$ and a radius of about $0.6~R_\odot$. 
It contributes about $4\%$ as much flux as the Be star does in the 
FUV.  We also present observations of the H$\alpha$ and \ion{He}{1} 
$\lambda 6678$ emission features that are formed in the 
circumstellar disk of the Be star.  Orbital flux and velocity 
variations in the \ion{He}{1} $\lambda 6678$ 
profile indicate that much of the emission forms along the 
disk rim facing the hot subdwarf where the disk is 
probably heated by the incident radiation from the subdwarf. 
A study of
the FUV infall shell lines discovered in the 1980s confirms 
their episodic presence but reveals that they tend to be found 
around both quadrature phases, unlike the pattern in Algol 
binaries. Phase-dependent variations in the UV \ion{N}{5} doublet 
suggest the presence of a N-enhanced wind from the subdwarf and a 
possible shock-interaction region between the stars where 
the subdwarf's wind collides with the disk of the Be star.

\end{abstract} 
 
 
 
\setcounter{footnote}{3} 
 
\section{Introduction}                              
 
The component stars of massive interacting binaries experience  
profound transformations over their lifetimes.  Their evolutionary  
stages are defined by how mass and angular momentum are shared and  
lost during the mass transfer stages \citep*{wel01}.  The originally 
more massive star grows faster and upon reaching a size comparable to  
its Roche lobe begins mass and angular momentum transfer to the  
mass gainer companion.  The orbit shrinks until the mass ratio reverses, 
but mass transfer will continue with an expanding orbit as long as  
the donor grows in proportion.  The result will be a donor that is  
stripped of most of its outer envelope and a gainer that now appears  
rejuvenated by mass transfer \citep{dra07}.  Mass transfer may  
also cause the gainer to spin up to a rotational speed close to the 
critical value where centripetal and gravitational accelerations  
are equal at the equator \citep*{pet05}.  \citet{pol91} and 
\citet{van97} suggest that these mass gainer stars may include 
some of the rapidly rotating, classical Be stars,  
B-type stars that display emission lines formed in outflowing,  
circumstellar disks \citep{por03}. 
 
There is now a substantial body of evidence that the supports  
the idea that some Be stars were spun up by mass transfer in interacting  
binaries.  Most of the known massive X-ray binaries consist of a Be star  
and neutron star companion, the remains of the former donor star \citep{coe00}. 
Donor stars that end up with a mass lower than the Chandrasekhar limit 
will probably avoid a supernova explosion, but will appear first as a  
hot subdwarf star and then eventually as a massive white dwarf.  Although  
there are no known examples of Be plus white dwarf binaries, there are  
a few identified cases of Be plus hot subdwarf binaries.  The first  
discovery of the spectrum of a subdwarf companion was made from the 
analysis of UV spectra from the   
{\it International Ultraviolet Explorer (IUE)} spacecraft \citep{tha95}  
and the {\it Hubble Space Telescope}  
\citep{gie98} of the well-known Be star $\phi$~Persei.  The hot companion in  
\object{$\phi$~Per} is an object of solar mass and radius but with an effective  
temperature of $53$~kK.  However, the hot subdwarf is much fainter  
than the Be star, and its presence is perhaps best observed through its heating  
effects on the disk gas nearest to the subdwarf \citep*{ste00,hum01}.  
\citet{mai05} show that the binary 59~Cygni shows many of the same  
characteristics found in $\phi$~Per and is probably the second example 
of a Be plus hot subdwarf system.  A third candidate, 
\object{HR~2142} \citep{pet83}, may also  
belong this category \citep{wat91} but at the present the evidence is 
not compelling.  The fact that the nature of the companions in $\phi$~Per 
and \object{59~Cyg} were only discovered after a century of spectroscopic 
observation, however, highlights the difficulties 
of their detection \citep{gie00}.  
 
Here we report on the identification of a hot subdwarf companion in the Be 
binary \object{FY~CMa} (HR 2855, HD 58978).  Balmer emission lines in the 
Be star primary in this system (B0.5~IVnpe,  
\citealt*{hil69}; B0.5~IVe, \citealt*{sle82}) were first discovered 
by \citet{pic05}.
Although at first glance it appears to be a classical Be star,
it has historically displayed much more spectroscopic variability
in its optical and UV spectrum than is typical for Be stars
\citep{sle82,pet82,pet88,gra88,cao01}.  Some of the first  
{\it IUE} observations of FY~CMa showed evidence of sharp, red-shifted, 
``shell'' spectral lines that varied significantly on a timescale of days  
\citep{gra88}.  \citet{pet88} argued that the time scales of  
variability indicated a binary origin. Spectroscopic variability with a  
period of 37.26~d was subsequently found by \citet{riv04} who showed that 
the strength and profile of the \ion{N}{5} $\lambda 1240$  
wind line fluctuates with this period.  \citet{riv04} also  
presented observations of the H$\alpha$ and \ion{He}{1} $\lambda6678$  
emission lines that suggested that they also varied in radial velocity  
and strength in much the same way as observed in $\phi$~Per and 59~Cyg,
and suggested that FY~CMa might be generically similar to these systems.  
However, \citet{riv04} could not confirm binarity as the orbital 
elements and the nature of the companion remained unknown. 
 
Below we present an investigation of FY~CMa that is based upon 
a re-examination of the {\it IUE} spectra and  
an analysis of a large collection of optical spectra obtained between  
1985 and 2006 (\S2).  We first discuss the use of cross-correlation  
measurements of the {\it IUE} spectra to obtain radial velocities for  
both components and to derive the orbital elements (\S3).  We then  
use the orbital velocity curves to reconstruct the individual UV spectra  
of the components using a Doppler tomography algorithm, and we compare 
the reconstructed spectra to models to estimate basic stellar properties (\S4). 
We then describe the orbital and temporal variations we find in the  
H$\alpha$ and \ion{He}{1} $\lambda6678$ emission lines (\S5).  We argue 
in \S6 that the emission line variations are probably due to heating  
of the part of the disk rim that faces the hot subdwarf companion.
We also discuss the long-term and phase-dependent variability in the 
strength and velocity of the \ion{Fe}{3} shell-type infall lines that were 
discovered in the mid-1980s and suggest two possible mechanisms for their 
formation. Finally, we show that phase-dependent profile, strength, and 
velocity variability in the \ion{N}{5} wind lines suggest that
the subdwarf's wind is N-enhanced and that there is a significant 
interaction between the subdwarf's wind and the disk of the Be 
star.   
 
 
\section{Observations}                              
 
Ninety-seven high resolution FUV spectra were obtained with {\it IUE} 
over the period from 1979 to 1995.  
The spectra were secured with the Short Wavelength Prime (SWP) camera 
in echelle mode, and the extracted and flux calibrated  
spectra from the final archive were downloaded from the Multimission Archive at 
the Space Telescope Science Institute\footnote{http://archive.stsci.edu/iue/}. 
We binned each spectrum onto a uniform $\log\lambda$ wavelength grid  
(bin size equivalent to 10 km~s$^{-1}$),  
aligned them with a global average using cross--correlation 
shifts from the vicinity of strong interstellar lines (after which the 
interstellar lines are removed by interpolation), and then rectified the 
fluxes by a spline fit to a series of pseudo-continuum zones.   
We discuss the radial velocity measurements and the appearance of the  
UV spectra in the next two sections.  
 
We also obtained 158 red spectra of FY~CMa with the Kitt Peak  
National Observatory 0.9~m Coud\'{e} Feed Telescope between 1985 and  
2006.  The various observing runs are summarized in Table~1 that 
lists the heliocentric Julian dates of observation, 
Besselian year range, wavelength range, 
spectral resolving power, number of individual spectra 
obtained, and a summary of the instrumental configuration. 
The spectra all record the H$\alpha$ $\lambda 6563$ emission line  
and all but six also record the \ion{He}{1} $\lambda 6678$ feature.  
All the spectra were made with the long collimator, camera 5,  
and either grating A (632 grooves mm$^{-1}$) or  
grating B (312 grooves mm$^{-1}$) in second order.   
The CCD detector of choice changed over the years  
as better instruments became available, and these are  
listed in the final column using their standard KPNO  
designations.  Most of the spectra are well exposed 
and have a S/N of 200 or better in the continuum. 
 
\placetable{tab1}      
 
The spectra were extracted and calibrated 
using standard routines in IRAF\footnote{IRAF is distributed by the 
National Optical Astronomy Observatory, which is operated by 
the Association of Universities for Research in Astronomy, Inc., 
under cooperative agreement with the National Science Foundation.}. 
All the spectra were rectified to a unit continuum by fitting 
line-free regions with a low-order polynomial.  During most of the runs  
from 2000 to 2006, we also observed a number of rapidly rotating  
A-type stars (primarily $\zeta$~Aql) for the removal of  
atmospheric telluric lines from the target spectra.  We  
created a library of atmospheric spectra for each run by removing 
the broad stellar features, and then we divided each target 
spectrum by the modified atmospheric spectrum that most closely 
matched the target spectrum in a selected region dominated by 
atmospheric absorptions.  The spectra from runs before 2000 were  
transformed to the wavelength grid for the fourth to last run listed  
in Table~1 and then were corrected for telluric absorption using the atmospheric 
spectra from that run (sometimes with less than optimal results because 
of differences in spectral resolution).  The spectra from each run were then 
transformed to a common heliocentric wavelength grid.  
Measurements of these red spectra are presented in \S5. 
 
 
\section{Radial Velocities and Orbital Elements}    
 
The first detection of the UV spectrum of the hot subdwarf in  
$\phi$~Per was based upon {\it IUE} spectra that recorded many  
lines of high excitation transitions especially in the 1310 -- 1385 \AA 
~range \citep{tha95}.  Thus, we began our search for evidence of  
a hot companion by calculating cross-correlation functions (ccfs) of  
each spectrum with a template spectrum within this restricted  
wavelength range (where the lines of the primary are relatively  
broad and weak).  We formed the template spectrum from the synthetic  
spectral atlas associated with the non-LTE, line-blanketed models of  
\citet{lan03} that were rebinned to the {\it IUE} wavelength grid. 
We selected a solar abundance model with $T_{\rm eff}=50$~kK and  
$\log g = 4.5$ for the template (close to the optimal values; \S4). 
The resulting ccfs immediately revealed a narrow but  
weak signal that displayed a coherent Doppler shift  
when plotted against the orbital period from \citet{riv04} (see Fig.~1).  
We then made Gaussian fits of these ccfs to determine the radial  
velocity shifts for 50 of the 97 spectra where the ccf signal was  
clearly visible.   We used the method of \citet{zuc03} to estimate 
errors in the ccf velocities, and these range from 7 to 20 km~s$^{-1}$
depending on the S/N ratio of the spectrum.

\begin{figure}
\begin{center} 
{\includegraphics[height=16cm]{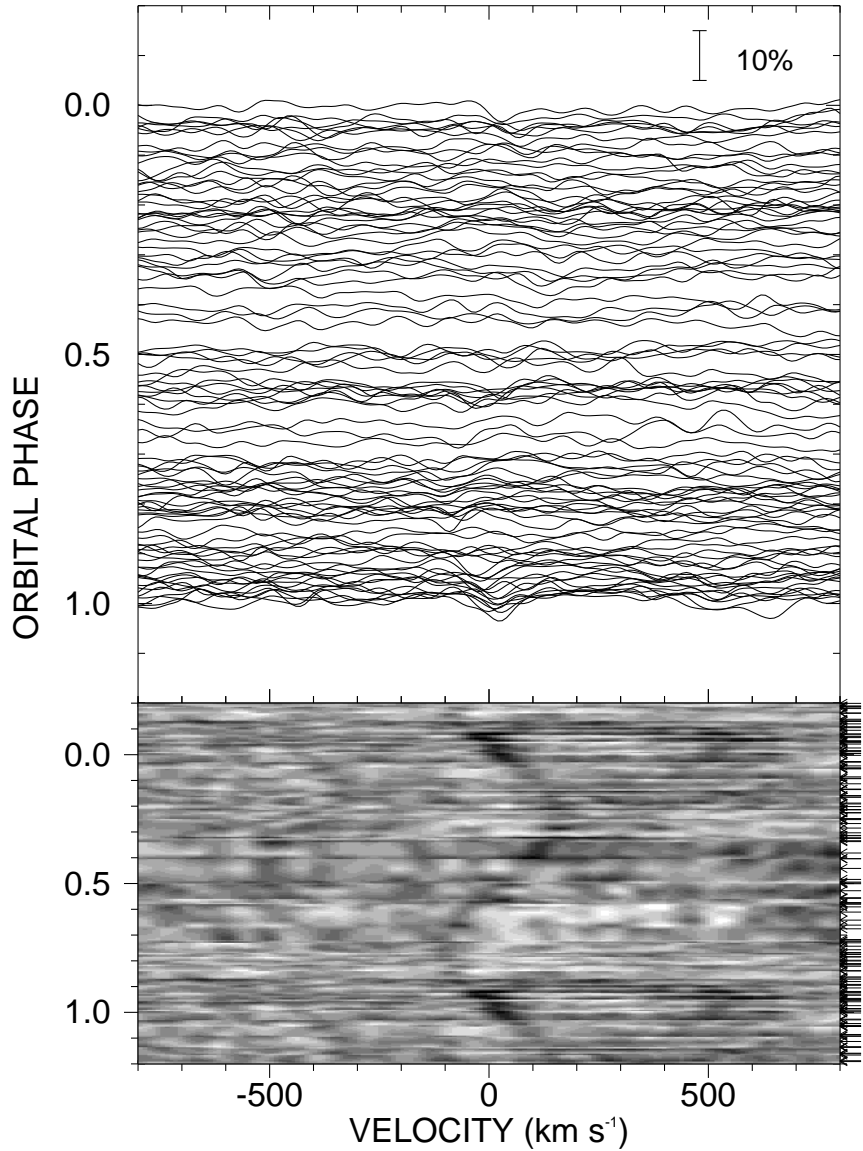}}
\end{center} 
\caption{The orbital phase variations in the cross-correlation 
functions of the spectra in the $1310-1385$ \AA ~region with 
a hot star template spectrum.  The ccfs are shown as 
linear plots ({\it top panel}) and as a gray-scale image ({\it lower panel}).  
The intensity in the gray-scale image is assigned one of 16 gray levels  
based on its value between the minimum (dark) and maximum (bright) observed values.  
The intensity between observed ccfs is calculated by a linear interpolation  
between the closest observed phases (shown by arrows along the right axis). 
The weak signal from the hot subdwarf spectrum appears as a backwards ``S''
feature in the lower panel.  The scale bar at top right indicates the 
amplitude of the ccf variation relative to the mean sum of the 
squared differences far away from the optimal correlation shift. 
\label{fig1}} 
\end{figure}

\placefigure{fig1}     
 
We also measured ccf radial velocities for the Be star primary  
using two spectral regions where its photospheric lines are  
particularly strong (1423 -- 1523 and 1570 -- 1800 \AA ). 
In this case we used a template spectrum derived from the  
models of \citet{lan03} for $T_{\rm eff}=30$~kK and $\log g = 4.0$, 
first guesses based upon calibrations for the star's spectral  
classification.  The resulting ccfs are broad because of the  
star's large projected rotational velocity, and instead of  
fitting these with a Gaussian, we determined the radial velocity  
shifts relative to the average ccf. According to the method
of \citet{zuc03}, the errors in the ccf velocities for the Be star
are about 3~km~s$^{-1}$. We then added the Gaussian-fitted 
radial velocity of the average ccf ($V_r = 34.3\pm 0.3$ km~s$^{-1}$) 
to each of these relative velocities to end up with the primary  
star velocities listed in Table~2.  This table lists the heliocentric
Julian and UT dates of mid-exposure, the {\it IUE} image number 
for the spectrum, the orbital phase (see below), and
the radial velocities $V_1$ and $V_2$ for the Be star (subscript 1) and
the hot subdwarf (subscript 2), respectively.  Also given with the same
subscripts are the observed minus calculated residuals $(O-C)$ from the
orbital fits (see Table~3), and we see that in general the
deviations from the fit are comparable to the error estimates
and to the scatter between measurements of pairs of spectra obtained
close in time.   
 
\placetable{tab2}      
 
We made a period search of the two sets of velocities using 
the discrete Fourier transform methods of \citet*{rob87}, and this  
analysis confirmed the period found by \citet{riv04}. 
We then used the non-linear, least-squares program of  
\citet{mor74} to determine independent orbital elements for  
the primary and secondary.  These produced consistent results  
for the orbital period and epoch, and we adopted the period and  
epoch from the larger amplitude secondary that had significantly 
lower errors.  We found that elliptical solutions did not yield  
any significant improvement in the residuals to the fits, so we 
assumed that the orbit is circular.  
Our final orbital elements are presented in column 3 of Table~3  
and the radial velocity curves are plotted in Figure~2.   
Table~3 lists period, $P$, epoch of the superior  
conjunction of the Be star, $T_{SC}$ (defining orbital phase 0.0),
semiamplitudes, $K$, and systemic velocities, $\gamma$, for
both components, mass ratio, $M_2/M_1$, mass--inclination functions,  
$M\sin ^3 i$, projected semimajor axis, $a\sin i$,
and r.m.s.\ residuals for each fit (where subscripts 1 and 2 
refer to the orbits of the Be and subdwarf star, respectively)  
 
\placetable{tab3}      
 
\begin{figure} 
\begin{center} 

{\includegraphics[angle=90,height=12cm]{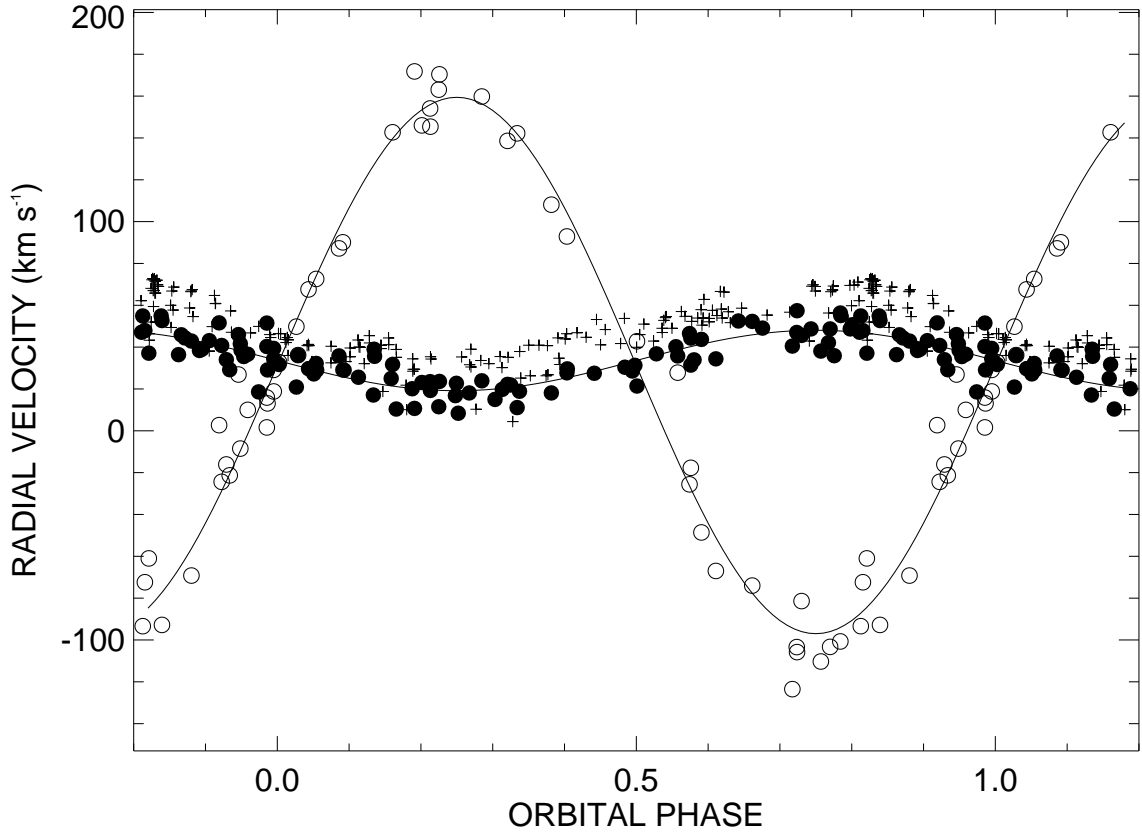}} 
\end{center} 
\caption{Radial velocity curves for the Be star and its companion. 
Orbital phase 0.0 corresponds to the Be star superior conjunction time.  
The solid circles represent the {\it IUE} cross-correlation  
velocities for the Be star while the open circles show the  
{\it IUE} cross-correlation results for the hot subdwarf.  
The plus signs indicate the velocities from measurements of the  
H$\alpha$ wing bisector positions. The typical radial velocity errors are 
3 and 7 km~s$^{-1}$ for measurements of the Be star and hot subdwarf, 
respectively.
\label{fig2}} 
\end{figure}

\placefigure{fig2}     
 
 
\section{Tomographic Reconstruction of UV Spectra}  
 
We used the Doppler tomography algorithm described by  
\citet{bag94} to reconstruct the individual primary and  
secondary spectra from the composite {\it IUE} spectra.  
We took the radial velocity shifts for each component  
from the orbital solutions in Table~3, and then the reconstruction  
was run for 50 iterations with a gain of 0.8 (the results  
are insensitive to these values).  The flux ratio required for  
the reconstruction was determined from the secondary's line  
depths after making a comparison of the secondary's spectral properties 
with those in model spectra.  
 
We began our comparison of the reconstructed secondary spectrum  
and model spectra from \citet{lan03} by considering the  
appropriate line broadening.  We selected several strong and  
minimally blended metallic lines to compare with a grid of  
synthetic profiles formed by convolution with a rotational broadening  
function for a linear limb darkening law \citep{wad85} and  
a range in assumed projected rotational velocity $V\sin i$.  
The resulting mean of $V\sin i = 41 \pm 5$ km~s$^{-1}$ indicates  
that the secondary's lines are narrow and only barely resolved in the  
{\it IUE} spectra.  We then experimented with a range of model  
effective temperatures and gravities to try to reproduce  
the relative patterns of line strength (especially in the  
region surrounding the weak \ion{Si}{3} $\lambda 1299$ feature). 
We also compared the reconstructed and model profiles of  
the Stark broadened \ion{He}{2} $\lambda 1640$ feature to  
help estimate the appropriate value of gravity $\log g$.  
The parameters for the best matching solar abundance model  
spectrum are listed in column 3 of Table~4.   
 
\placetable{tab4}      
 
We then estimated the mean UV flux ratio to be  
$F_2/F_1 = 0.04 \pm 0.01$ based upon a comparison of the  
line strengths in the best match model and in the reconstructed  
secondary spectrum for a range in test values of flux ratio.  
We show in Figures 3 and 4 the good agreement between the  
reconstructed secondary and model spectra for two regions  
of the UV spectrum that contain a number of prominent  
\ion{Fe}{5} and \ion{O}{5} photospheric lines.  
 
\begin{figure} 
{\includegraphics[angle=90,height=12cm]{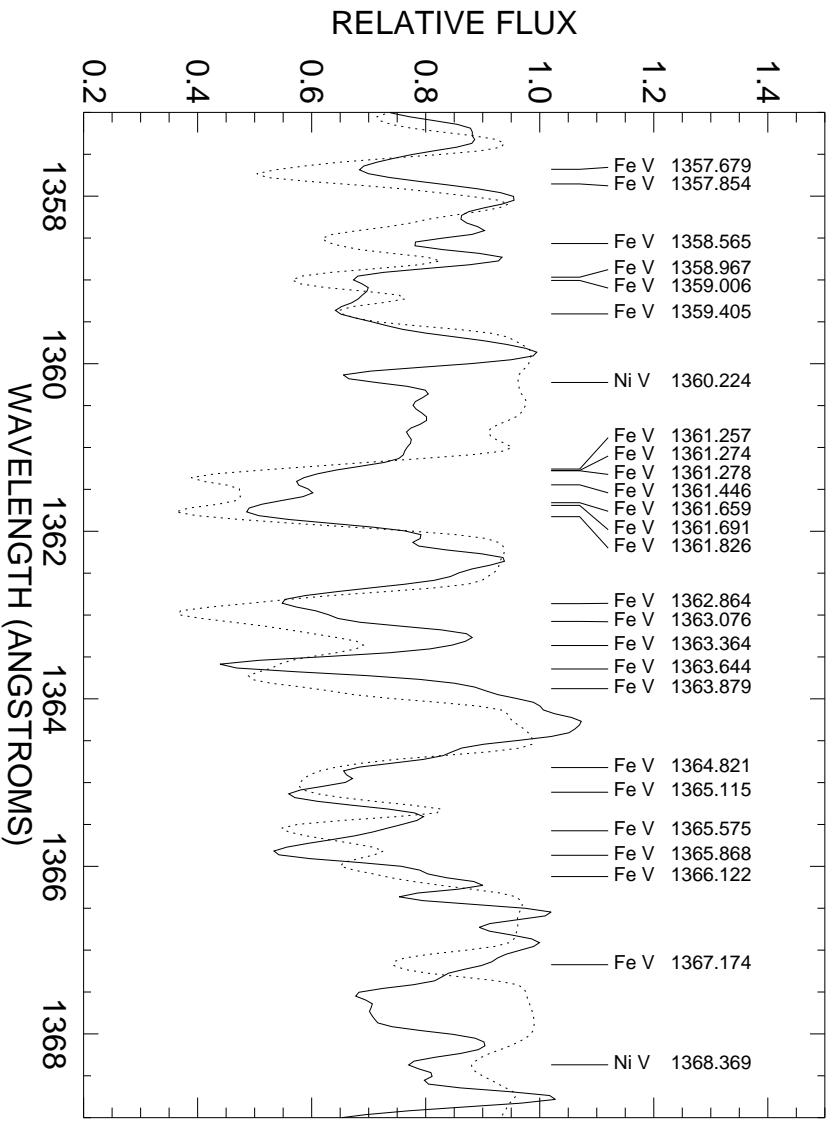}} 
\caption{A comparison of a part of the reconstructed UV spectrum of the 
secondary star ({\it solid line}) with a model spectrum ({\it dotted line}). 
Line identifications for some of the stronger lines are given above the spectra. 
\label{fig3}} 
\end{figure}

\placefigure{fig3}     
 
\begin{figure} 
{\includegraphics[angle=90,height=12cm]{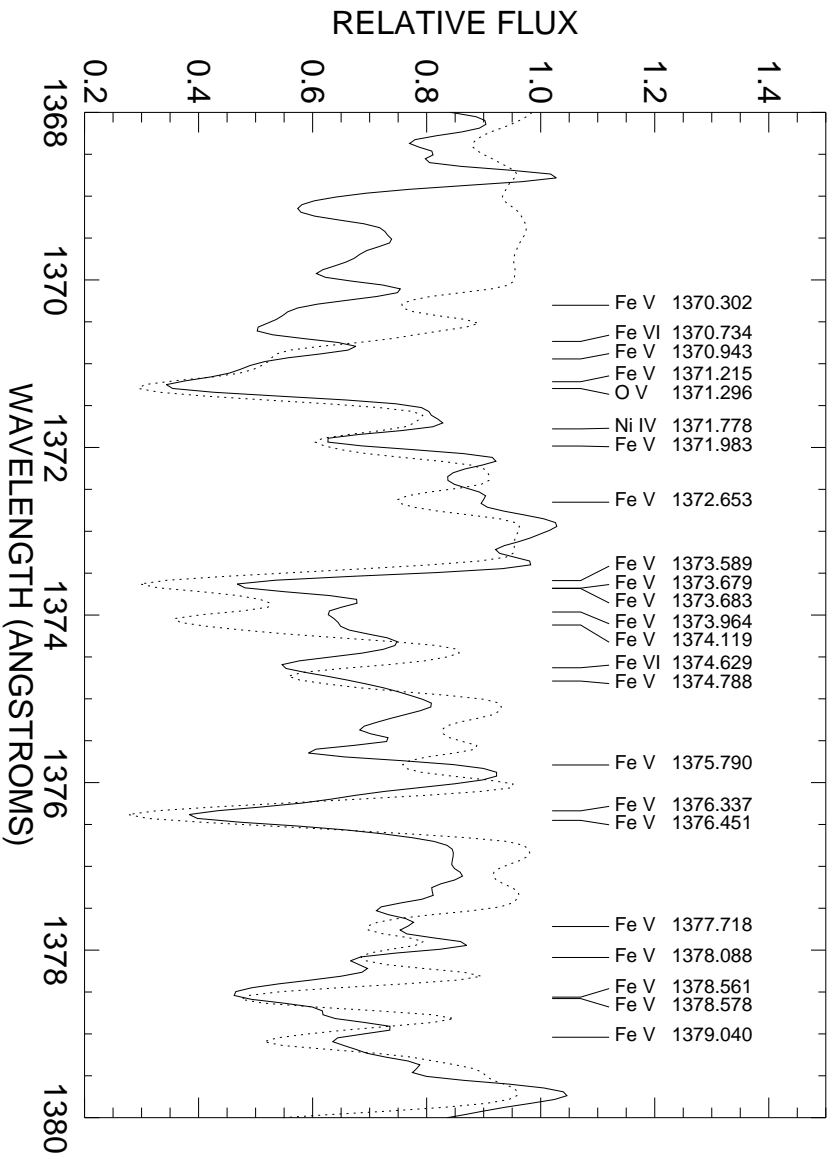}} 
\caption{A comparison of another part of the reconstructed UV spectrum of the 
secondary star ({\it solid line}) with a model spectrum ({\it dotted line}). 
\label{fig4}} 
\end{figure}

\placefigure{fig4}     
 
We also explored the range in synthetic spectral matches  
that were consistent with the appearance of the reconstructed 
Be star spectrum.  Once again we compared the profile widths 
of several conspicuous absorption lines with models for  
a range in projected rotational velocity, and we found  
a mean value of $V\sin i = 340 \pm 40$ km~s$^{-1}$.  
This estimate is similar to the result from \citet{fre05},
375 km~s$^{-1}$, but it is larger than those reported  
in other optical studies (280 km~s$^{-1}$, \citealt{sle82}; 
155 km~s$^{-1}$, \citealt*{abt02}).  We suspect that 
the optical lines may at times appear narrower  
because of unrecognized shell components formed in  
the Be star's disk.  We estimated the best match effective 
temperature based on the metallic line strengths  
(avoiding the use of narrow features like \ion{Si}{3} $\lambda 1299$ 
that are probably partially formed in the disk; \S6), 
and the best match model has a temperature consistent with 
the star's spectral classification (Table~4).  We simply 
assumed the gravity for the Be star based on its luminosity class 
since there are no easily made gravity estimates from the 
UV photospheric lines.  We note that both the temperature
and the gravity agree well with estimates made by \citet{pet76} 
based upon the continuum colors and the profiles of 
H$\gamma$ and H$\delta$.
 
FY~CMa is not a known eclipsing binary, so we cannot directly  
determine the system inclination and masses.  However, the appearance 
of shell lines in the UV and optical spectra suggests that the 
inclination is probably large, $i > 77^\circ$ \citep{han96}, if the  
disk resides in the orbital plane.  The typical mass for a B0.5~V star 
is $\approx 13.2 M_\odot$ \citep{har88}, and from the value of $M_1 \sin ^3 i$ 
in Table~3, this would imply that the inclination could be as small  
as $i=66^\circ$.  The range in probable masses is given in Table~4. 
 
We can estimate the radius of the Be star primary from its visual  
magnitude $V$ and its parallax $\pi$ as measured by {\it Hipparcos} \citep{esa97}.
Adopting $V=5.63$ \citep*{ste96}, $E(B-V)=0.14$ \citep{kai89}, 
$A_V=3.1 E(B-V)$, $\pi = 2.30\pm 0.70$ mas, and a bolometric correction  
of $BC=-2.65$ \citep{lan03}, then the radius from the luminosity  
and adopted temperature is $R_1 = 5.3 R_\odot$, where we  
have ignored the minor flux contributions from the disk and subdwarf.  
The radius of the subdwarf then follows from the observed and model  
UV flux ratios, $R_2 = 0.63 R_\odot$.  These mass and radius  
estimates yield gravities that are similar within errors to those  
from spectroscopic considerations.   
 
 
\section{Red Spectra Measurements}                  
 
The H$\alpha$ feature in our spectra always appeared as a 
strong emission line indicating that the Be disk was present  
throughout the period 1985 -- 2006.  We show in Figure~5 a montage  
of all the H$\alpha$ profiles arranged according to orbital phase.  
The profile is often double-peaked, usually in the sense that  
the red peak is stronger around phase $\phi=0.25$ while the  
blue peak is stronger near phase $\phi=0.75$.  Thus, the  
observations appear to confirm the suggestion by \citet{riv04}  
that the variations are orbitally modulated.  On the other hand,  
our spectra also show that there are temporal variations present 
unrelated to orbital phase that appear as alternating bright and 
dark horizontal segments in the gray-scale representation.  
We measured the H$\alpha$ equivalent widths by making a  
numerical integration of the line flux over the interval  
from 6528 to 6599 \AA ~(a region including weak emission  
components of \ion{C}{2} $\lambda 6578,6582$), and these 
are listed in Table~5. The typical equivalent width measurement 
error is 0.04 \AA . We find that there are significant variations 
in line strength over time scales of a week or so that appear to be 
unrelated to orbital phase and that probably reflect changes in the 
disk density distribution. 
 
\begin{figure}
\begin{center} 
{\includegraphics[height=16cm]{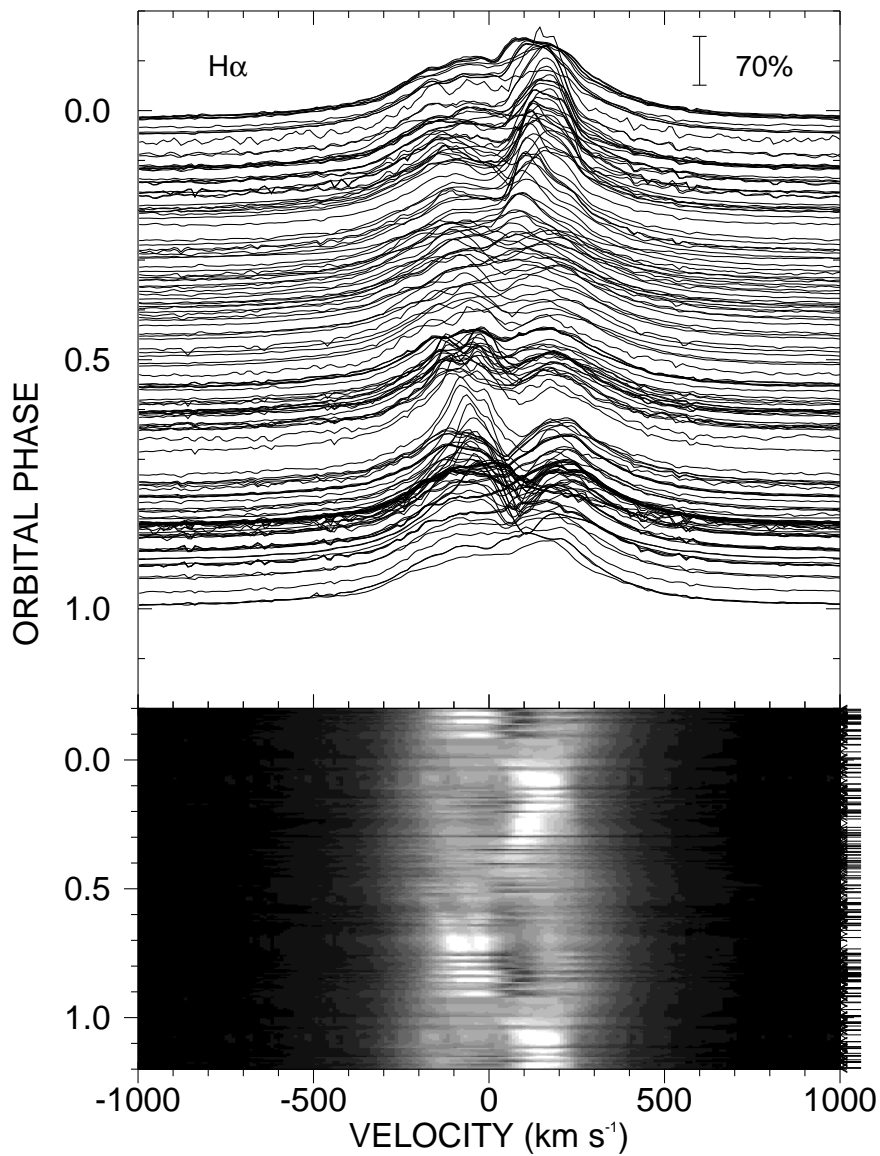}}
\end{center} 
\caption{The orbital phase variations in the H$\alpha$ 
emission line in the spectra of FY~CMa are shown in  
linear plots ({\it top panel}) and as a gray-scale image ({\it lower panel}).  
The intensity in the gray-scale image is assigned one of 16 gray levels  
based on its value between the minimum (dark) and maximum (bright) observed values.  
The intensity between observed spectra is calculated by a linear interpolation  
between the closest observed phases (shown by arrows along the right axis). 
The scale bar at top right indicates the spectral flux 
relative to the local continuum flux.
\label{fig5}} 
\end{figure}

\placefigure{fig5}     
 
\placetable{tab5}      
 
The outer wings of H$\alpha$ show a hint of an ``S''-shaped  
velocity variation that is similar to the orbital velocity  
curve of the Be star.  Since the outer parts of the profile  
are formed in the disk gas closest to the star, the line wings may  
share in the orbital motion of the underlying star.  We measured  
the bisector position of the emission line wings near the  
25\% maximum flux level using the method of \citet*{sha86},  
and these wing velocities are listed in column 4 of Table~5 
and are plotted together with the UV ccf velocities in Figure~2.
The typical error for the radial 
velocities from the H$\alpha$ wings is 1.0 km~s$^{-1}$.   
The H$\alpha$ wing velocities indeed follow the orbital  
motion of the Be star, and the orbital elements based  
upon them are in substantial agreement with those from  
the UV ccfs (see column 2 of Table 2).  The differences in  
systemic velocity $\gamma$ are not significant given the asymmetries 
and line blending present in the H$\alpha$ profile.  
 
We show the dramatic variations of the \ion{He}{1} $\lambda 6678$  
profile with orbital phase in Figure~6.  Here the emission displays  
a backwards ``S''-shaped velocity variation that is similar to
the secondary's orbital velocity curve.  The emission is  
strongest around orbital phase 0.0 when the hot subdwarf is in  
the foreground.  The emission sometimes assumes a pronounced  
double-peaked shape, especially when the emission is relatively  
strong.  In the same way as H$\alpha$, the \ion{He}{1} line also 
displays non-orbital related variations that give a banded  
appearance to the gray-scale representation of the variations.  
We also see the occasional development of a central  
shell absorption feature at phases prior to superior conjunction  
of the Be star. 
 
\begin{figure} 
\begin{center} 
{\includegraphics[height=16cm]{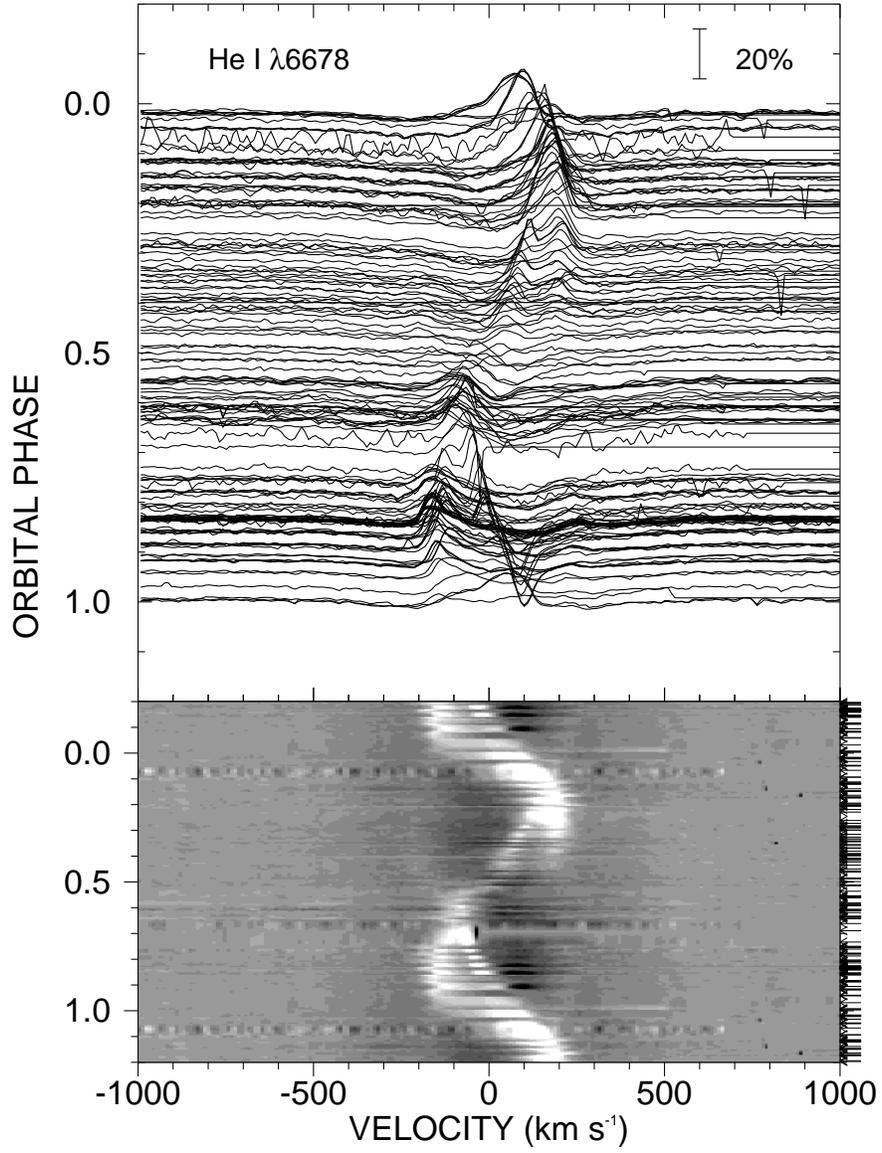}} 
\end{center} 
\caption{The orbital phase variations in the \ion{He}{1} $\lambda 6678$  
emission line plotted in the same format as Fig.~5. 
\label{fig6}} 
\end{figure}

\placefigure{fig6}     
 
We made simple measurements of the peak velocity positions by  
locating the maximum flux positions in the \ion{He}{1} profile. 
If two peaks were present and clearly separated by a local  
minimum, then we measured both peaks and recorded their  
mean position $V_r$ and their separation $\triangle V_r$. 
If only one peak was visible, then only $V_r$ was measured.  
These measurements are given in the final two columns of  
Table~5 for all those cases where at least one strong peak  
was visible.  They are also plotted in Figure~7 together  
with the orbital velocity curve of the hot subdwarf.  
We see that the velocity amplitude of the \ion{He}{1} emission  
is actually larger than that of the subdwarf and that  
the episodes of greatest peak separation occur just prior  
to the orbital conjunctions.  We show in the next section 
that these characteristics are consistent with an emission  
origin in the outer part of the disk that faces the subdwarf.  
 
\begin{figure} 
{\includegraphics[angle=90,height=12cm]{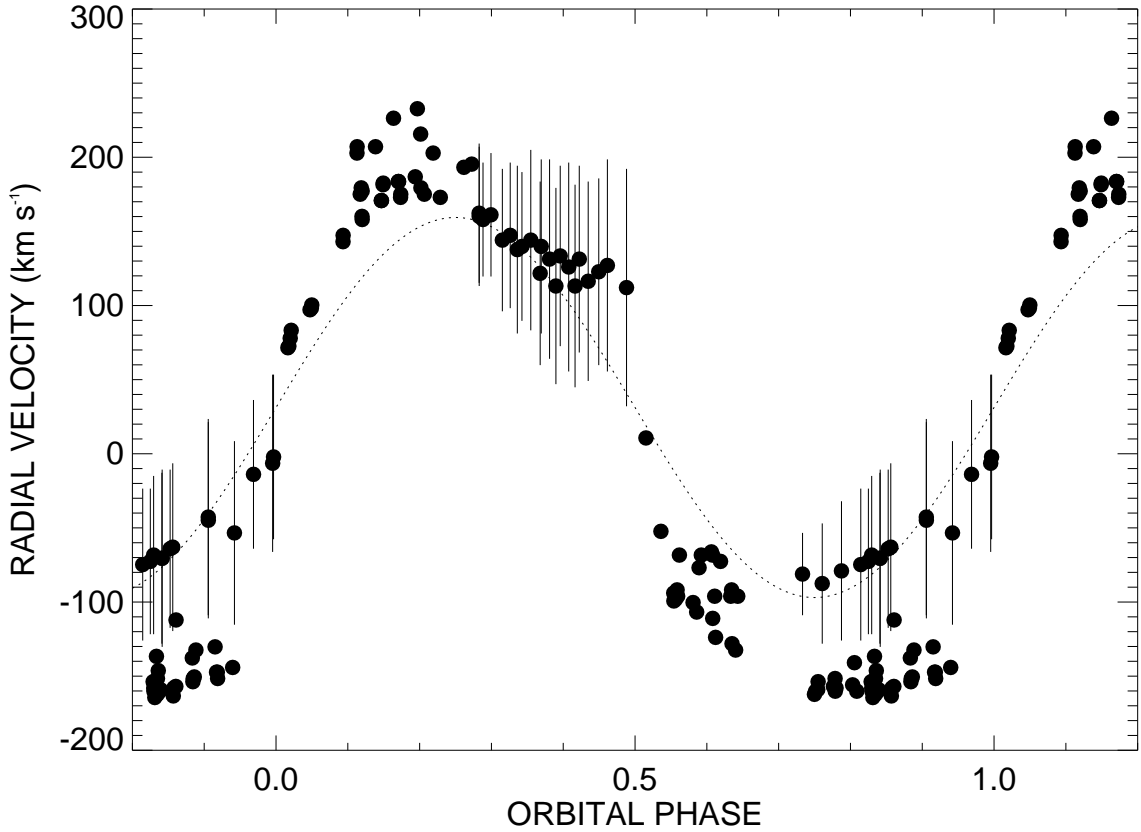}} 
\caption{The radial velocities of the prominent emission peaks in  
the profile of \ion{He}{1} $\lambda 6678$ plotted as  
a function of orbital phase.  Single dots represent measurements  
of one peak (possibly blended) while dots with vertical segments 
represent the midpoint velocity where two prominent peaks  
were observed.  The end points of these line segments correspond  
to the velocity of each resolved peak. The dotted line  
represents the orbital velocity curve of the hot subdwarf.  
\label{fig7}} 
\end{figure}

\placefigure{fig7}     
 
 
\section{Discussion}                                

Our spectroscopic results provide the ingredients to estimate
the basic elements of the system geometry.  Here we discuss 
the system dimensions and the evidence that the hot subdwarf 
is implicated in local heating of the Be star's disk in 
much the same way as was found for $\phi$~Per by \citet{ste00}
and \citet{hum01} and for 59~Cyg by \citet{mai05}. 
We present in Figure~8 a sketch of the inferred system geometry as 
viewed from above the orbital plane.  For the purpose of this
illustration, we assume an orbital inclination of $i=70^\circ$
and a semimajor axis of $112~R_\odot$ (where the acceptable 
range is probably $105 - 115 R_\odot$ for $i=90^\circ - 66^\circ$).
We also adopt a Be star radius of $R_1=5.3 R_\odot$ 
(where the probable range is $4.1 - 7.6 R_\odot$ for distances 
of $333 - 625$~pc). The Be star is represented by the solid disk
on the left while the hot subdwarf is shown by the small dot 
on the right.  The double-lobe surrounding them shows the 
much larger Roche surface boundary.   We can estimate the half light
radius of the H$\alpha$ emitting disk using the method described by 
\citet{gru06}.  This estimate is derived from a simple density power
law for the disk, the mean equivalent width ($W_\lambda = -13.6$ \AA ),
the adopted Be star temperature and disk inclination, and an assumed
disk outer boundary equal to the volume-equivalent, Roche radius.
These parameters lead to an estimated disk to star radius ratio of 
$R_d/R_1=9.4$, or a half-light emitting radius of $50 R_\odot$ for 
the adopted radius of the Be star.  
We caution that the simplicity of the models
may lead to errors in the half-light radius of $\pm 30\%$ and that the models 
assume an azimuthal symmetry for the emission flux, which is probably not the case 
for the disk emission of FY~CMa (see below).  Nevertheless, this simple estimate
of disk radius is indicated by the lighter gray shaded region on the left side of  
Figure~8.  \citet{oka01} argue that the disks of Be stars in binaries 
with nearly circular orbits are truncated at the 3:1 orbital resonance radius,
and this occurs at a distance from the Be star of $52 R_\odot$ that is only 
slightly larger than our disk half-light radius estimate. 
 
\begin{figure}
\begin{center} 
{\includegraphics[angle=90,height=12cm]{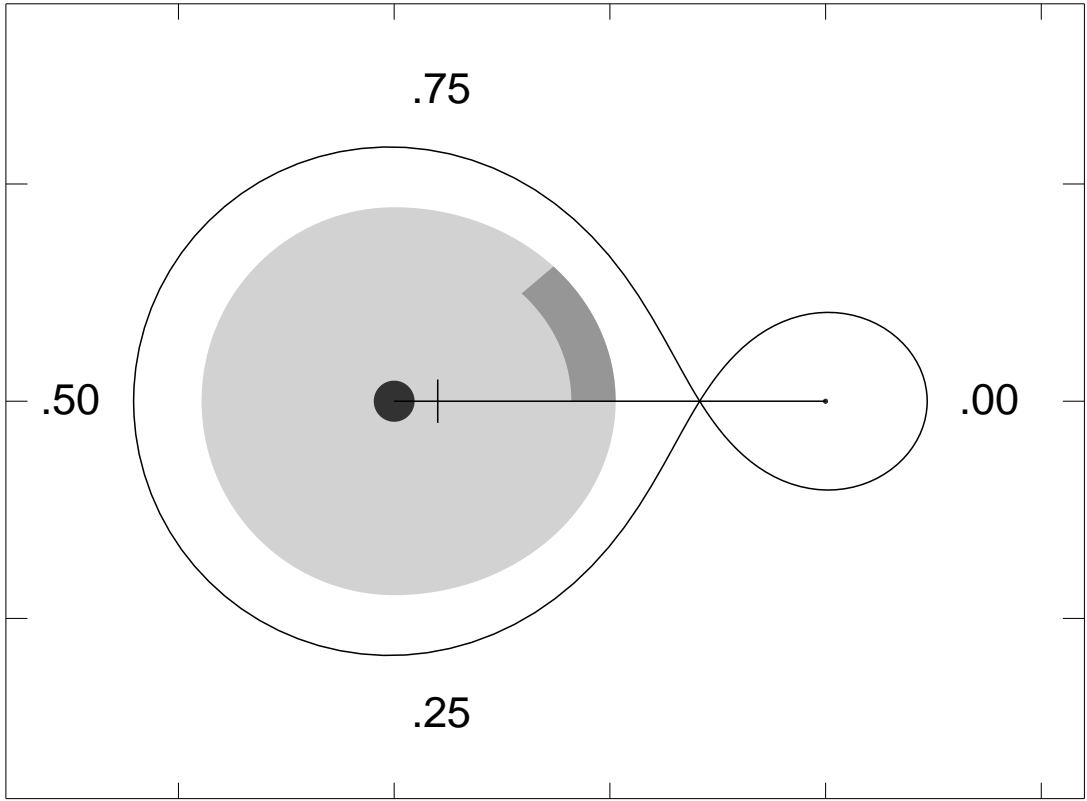}} 
\end{center}
\caption{An empirical model of the system geometry as viewed  
from above the orbital plane.  Each tick mark corresponds to a 
length of $0.5 a$. The Be primary ({\it left}) is surrounded  
by an equatorial disk ({\it lightly shaded region}) that is heated along  
the edge ({\it darker shaded region}) facing the subdwarf secondary ({\it right}).  
The outer boundary of the Be star disk is probably within the critical Roche surface 
({\it solid curve}). The center-of-mass of the system is indicated by the vertical tick
mark on the line-of-centers. The orbital phases are indicated on the periphery. 
\label{fig8}} 
\end{figure}

\placefigure{fig8}     

We showed in \S5 that the \ion{He}{1} $\lambda 6678$ emission has 
an orbital modulation that leads to greatest emission strength 
when the hot subdwarf is in the foreground.  Furthermore we found that the 
semiamplitude of the orbital Doppler shift of the emission peaks 
was larger than that of the subdwarf (from Fig.~7, $K_{em}\approx 173$
km~s$^{-1}$ $> K_2 = 128$ km~s$^{-1}$).  These facts suggest that a 
significant fraction of the emission forms in the part of the disk near the 
subdwarf where the Keplerian velocity of the disk gas is larger than the 
orbital velocity of the subdwarf. Ignoring three-body effects, we can 
estimate the radius of the emission source by assuming both
the disk gas and the subdwarf obey Keplerian motion around the Be star, 
$R_{em}/a = (M_1/(M_1+M_2)) ((K_1+K_2)/(K_1+K_{em}))^2 = 0.51$ 
or $R_{em}=58 R_\odot$.
This radius is about $15\%$ larger than the half-light radius,
which suggests that the emission forms mainly along the outer rim 
of a disk that may be extended towards the hot subdwarf (see Fig.~8).  
We suspect that this outer disk region is heated by the combined 
effects of the incident subdwarf radiation and the impact of the
subdwarf's wind on the disk.  We imagine that the rest of the disk also 
contributes to the line emission flux, and, for example, those times 
when weak emission is observed (see Fig.~7) may correspond to episodes 
when the gas density declines in the outer rim facing the subdwarf 
and the flux from the rest of the disk dominates.

Since the heated disk gas nearest the hot subdwarf will 
have a higher Keplerian velocity, the center of the heated region 
will move ahead of the axis joining the stars by an amount that 
depends on the cooling and orbital timescales.  We can estimate 
the phase offset $\triangle \phi$ of this asymmetry in two
ways.  First, the central heated region should reach inferior
conjunction about midway between the observed velocity extrema 
at phases $\phi=0.77$ and 0.17 (see Fig.~7), i.e., at $\phi=0.96$.
Second, the velocity peak separation is presumably caused by 
the heated region forming in a sector of a ring.  We will observe
the largest velocity separations when the mid-point of the ring segment 
is oriented along our line of sight.  According to the peak separations 
plotted in Figure~7, greatest peak separations occur around 
$\phi=0.42$ and $\phi=0.92$.   This suggests that the apex of the
heated rim is offset by about 0.06 in phase ($22^\circ$) ahead of the 
subdwarf inferior junction at $\phi=0.0$.  From simple geometric 
arguments, we expect that the maximum half-peak separation 
$V_m = {\rm max}(\triangle V_r /2)$ is related to the emission 
semiamplitude by $V_m=K_{em} \sin \theta$, where $\theta$ is 
half the ring sector opening angle.  From the values of 
peak separation in Table~5, we estimate $V_m=66$ km~s$^{-1}$ and 
$\theta = 22^\circ$.  These values were used to plot a 
representative heated region in the darker gray shaded area 
of Figure~8.  The actual extent of the heated ring is probably 
larger than shown since the separated emission peaks actually
sample a range of ring azimuthal angles and not just the termination
angles. 
 
The subdwarf may also affect the kinematics of the
gas motion in the disk rim.  We show in Figure~9 the orbital 
phase variations in the appearance of the \ion{Si}{3} lines 
near \ion{Si}{3} $\lambda 1299$.  The cores of these
lines are shell-type features formed in the disk gas seen in projection 
against the photosphere of the Be star.  We see evidence that 
the shell feature broadens from $\phi=0.5$ to $\phi=0.0$ and 
apparently shifts from a redshifted to a blueshifted position
at phases surrounding conjunction at $\phi=0.0$. 
We imagine that the motion of the outer disk gas is influenced by 
subdwarf as the disk gas passes and overtakes the star.  
The  gravitational attraction would cause acceleration and outward motion 
as the disk gas approaches the subdwarf, but the same forces would cause 
deceleration and inward motion following closest passage.
Thus, we would observe infalling (redshifted) disk gas before $\phi=0.0$ 
and then outward moving (blueshifted) gas after $\phi=0.0$. 
On the other hand, the disk gas projected against the Be star has
more nearly Keplerian and tangential motion at $\phi=0.5$ leading to  
the appearance of narrow shell lines at that phase. 
This kind of perturbation in the Keplerian motions may also partially
explain the appearance of a redshifted shell feature in the 
\ion{He}{1} $\lambda6678$ profiles near $\phi=0.9$ (Fig.~6), 
an orientation where the disk gas projected against the Be star  
would be infalling and relatively dense. 

\begin{figure} 
\begin{center} 
{\includegraphics[height=16cm]{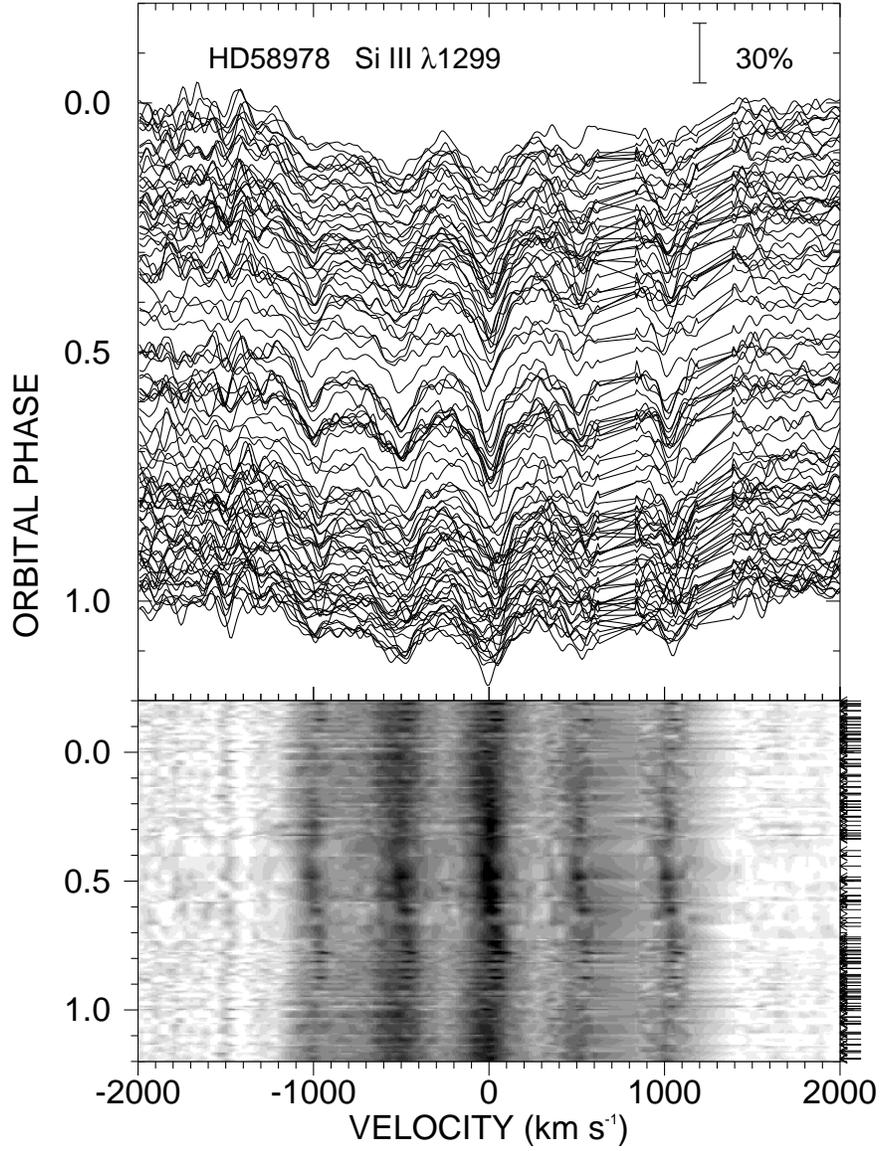}} 
\end{center} 
\caption{The orbital phase variations in \ion{Si}{3} $\lambda 1299$ 
and other nearby \ion{Si}{3} lines  
plotted in the same format as Fig.~5. These features are a composite 
of broad photospheric lines and narrower ``shell'' features formed in the disk. 
\label{fig9}} 
\end{figure}

\placefigure{fig9}     


In addition to the shell features discussed above that predominantly are 
formed in the disk, another circumstellar component is seen in some  
intermediately-ionized species (e.g. \ion{S}{3}, \ion{Fe}{3}). 
These variable features, which rarely show velocities 
less than the photospheric value and can display velocities as large as
50~$\rm km~s^{-1}$ relative to the photosphere of the Be star, were discovered 
in {\it IUE} images obtained in the 1980s \citep*{pet88,gra88} and they clearly 
show the presence of significant infall of material toward the Be star. Often 
times these lines are asymmetric with enhanced absorption on the red side 
of the profile. \citet*{gra88} identified six infall episodes that are best 
seen in the \ion{Fe}{3} (multiplet~34) line at 1895.456~\AA, since the feature 
does not have to compete with a strong photospheric component and it is not 
blended with interstellar or other circumstellar lines. 

From the entire set of 97 {\it IUE} SWP high resolution (HIRES) 
images that span more than 15 years of observation and our 
knowledge of the orbit of the system, we now have more insight on the
nature of the infall lines. Their behavior in strength and velocity are shown in
Figures~10, 11, and 12. Representative examples of the three basic types of 
observed features, photospheric, photospheric with additional absorption from the 
Be star's disk, and infall are given in Figure~10. In order to eliminate 
phase-dependent effects, we have chosen spectra taken at approximately the same 
quadrature phase on the {\it trailing} hemisphere of the Be star but at different 
epochs. Note that the central core of the profile taken during a period of enhanced 
disk absorption shows the same velocity as the photospheric line, but that the 
feature formed in infalling material has a significant redshift.
In the upper two-part panel of Figure~11 we plot the residual intensity\footnote{The
observed flux divided by the flux in the local continuum outside of the spectral 
feature.} of the 
deepest part of the \ion{Fe}{3} line and its radial velocity versus observation date.
All data are included in the intensity plot, but in the velocity plot we use a 
different symbol for observations that show no, or minimal shell absorption. Points 
that fall above the {\it dashed} horizontal line in the upper panel visually appeared 
to be mostly photospheric. The double horizontal {\it dashed lines}  in the second 
panel delineate the domain of photospheric motion. Features that fall above the 
double line are clearly associated with infall phases. Seven or eight clear infall 
episodes are observed. The data suggest that the duration of the infall phases varies 
from days to a year, but significant data gaps from 1979-85 and 1990-94 render it 
impossible to determine a firm value or its degree of variability. An apparently 
extended infall phase of about 300 days was seen from HJD~2446919--2447215. In 
the lower two panels of Figure~11, the same data are plotted versus orbital phase.
From the third panel it can be seen that on the average the shell-type absorption is 
enhanced in the phase interval 0.3--0.7. But from the display in the fourth panel it 
is apparent that features seen around phase 0.5 tend to display the systemic velocity, 
implying they are formed in the disk of the Be star.
In the fourth panel core velocities from only the spectra 
showing enhanced shell absorption are compared with the radial velocity curve of  
the Be star. It is clear that infall features with velocities up 
to 50~$\rm km~s^{-1}$ relative to the photosphere are seen {\it at most phases}
with a minimal occurrence around phase 0.5. The infall features tend to cluster 
around the quadrature points at which their velocities are highest relative to the
Be star's photosphere. Many of the phase dependent strength and velocity variations 
discussed above can also be seen in the gray-scale plot presented in Figure~12. 

The episodic infall of material is fundamentally different from that observed in 
the Algol-type binaries \citep{pet01}, where mass motion toward a mass gainer 
is typically seen only between phases 0.75--0.95. In the case of FY~CMa, the infall
features predominantly cluster around the quadrature phases from 0.15--0.35 and 
0.65--0.90. The velocities of the shell features around phase 0.5 tend to be either 
photospheric or negative relative to the Be star\footnote{Evidence for material
outflow is often seen in interacting binaries of the Algol type \citep{pet01} and 
similar hydrodynamics might be occuring in this Be + subdwarf system.}. \citet{pet88} 
discussed various scenarios to explain the observed simultaneous presence of infall 
and outflow features in this system including a sudden accretion event, 
the possible presence of magnetic loops, or flow from a polar jet. 
\citet{cao01} reported that infall components in the \ion{N}{5} doublet 
were contemporaneously observed with those in \ion{He}{1} $\lambda5876$ and also
suggested that magnetic loops might be involved, but the phase-dependence of the infall
features does not support this speculation. The infall is more likely a consequence of an 
interaction between the wind from the subdwarf and the periphery of the massive disk 
of the Be star or disk perturbations due to the tidal force produced by the subdwarf.
The motion of a shock-interaction region would result from variations in the location of 
the interface where there is a balance between the
dynamic pressure of the subdwarf's wind and the gas pressure at the periphery 
of the Be star's disk.  It is not clear at this time whether it is the subdwarf or
the Be star that might be driving the variability though. Epochs when no enhanced shell 
absorption is observed would correspond to minimal wind-disk interaction.
Alternatively the episodic infall features could simply be 
a result of gas dynamics in the disk due to a tidal effect from the subdwarf. 
Hydrodynamical simulations\footnote{cf. 
http://harmas.arc.hokkai-s-u.ac.jp/\%7Eokazaki/BeX/sim/index.html} of the 
gas motion in a perturbed disk in a Be/X-Ray binary
\citep{oka02} reveal turbulence in the outer disk and even the formation of 
spiral structure. Similar disk variability could be operating in FY~CMa.
The details are probably complex but an interplay of such gravitationally related
forces could create time variable and sometimes coherent gas flows.
For example, the tidal deformation of the disk leads to an axial elongation 
that moves ahead of the motion binary.  Disk gas returning from the
extended regions would be observed to be moving inwards when
viewed at the quadrature phases.

\begin{figure} 
\begin{center} 
{\includegraphics[height=10cm]{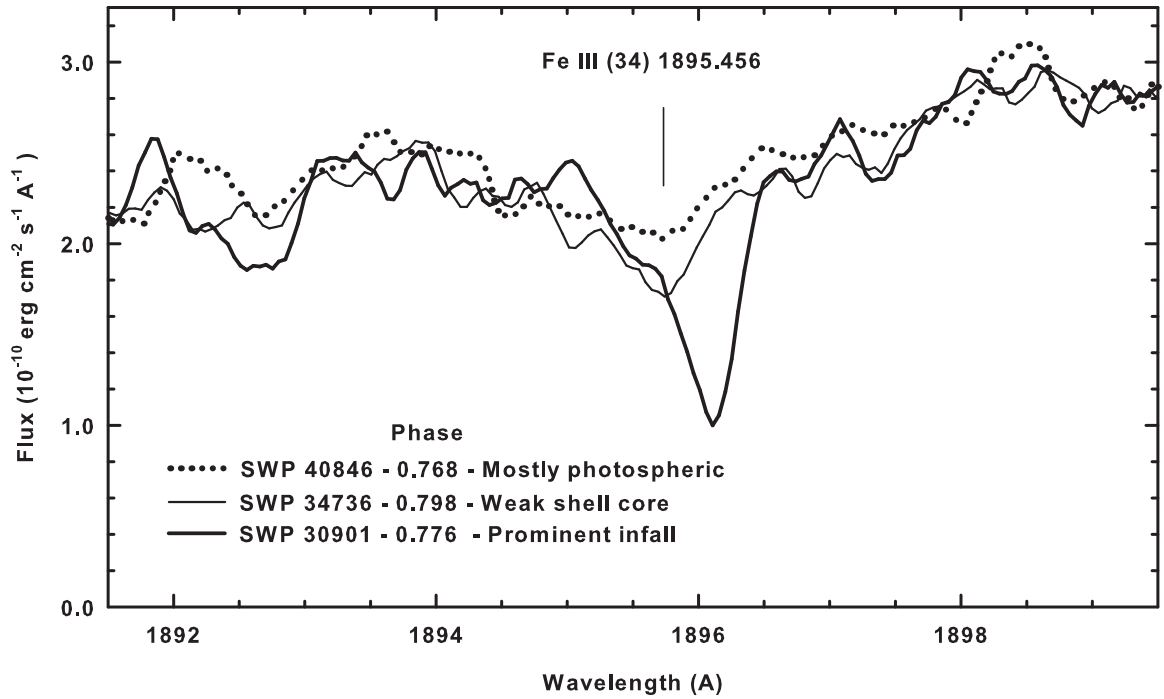}} 
\end{center} 
\caption{Selected profiles of the \ion{Fe}{3} $\lambda 1895$ feature, 
all observed near phase 0.75 but at different epochs. Examples include an 
apparently pure, rotationally-broadened photospheric profile ({\it dotted line}),
a profile with weak shell absorption ({\it thin, solid line}), and an observation
made during a prominent infall phase ({\it thick, solid line}). The vertical 
line above the profiles shows the observed line center that includes the velocity 
shift of the Be star. 
\label{fig10}} 
\end{figure}

\begin{figure} 
\begin{center} 
{\includegraphics[height=14.5cm]{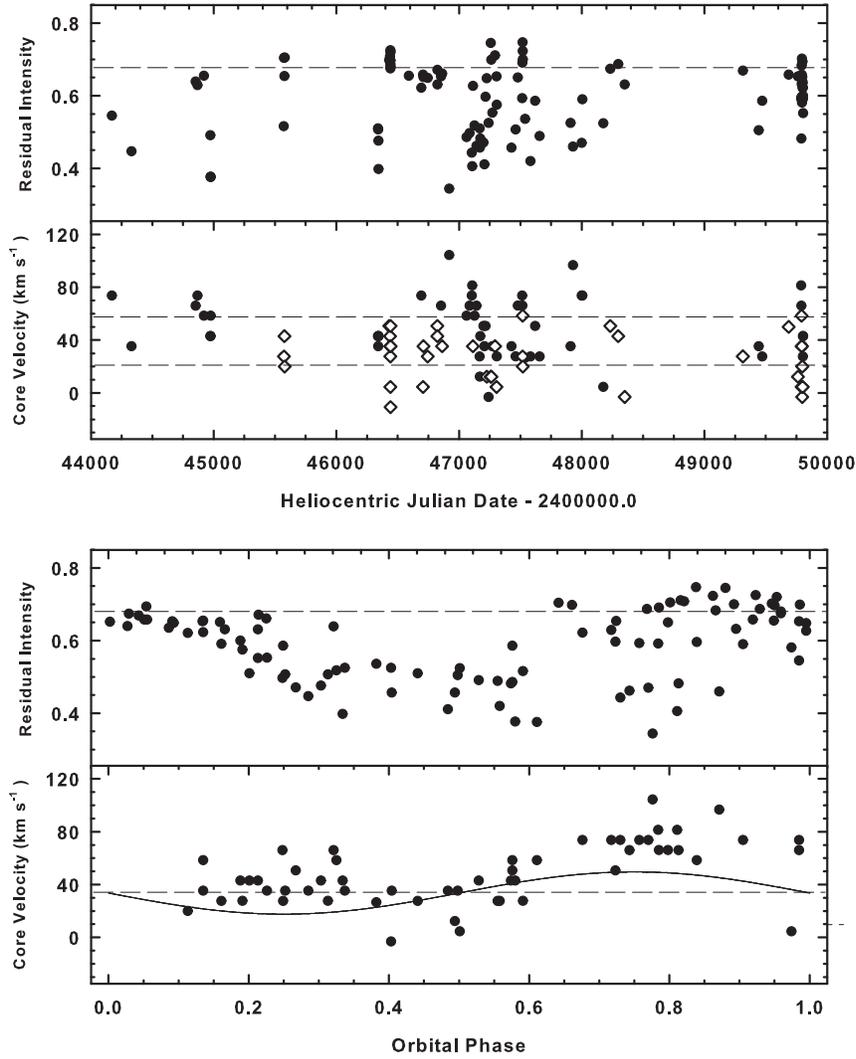}} 
\end{center} 
\caption{The absorption strength and velocity of the core of the \ion{Fe}{3}
$\lambda 1895$~\AA\, line versus time and orbital phase. 
\textit{Panel 1}: The horizontal {\it dashed line} represents the core intensity 
above which the feature appears to be primarily photospheric. 
\textit{Panel 2}: The double {\it dashed lines} represent the range in photospheric 
velocity of the Be star. The observations shown with the {\it open diamonds} 
showed either no or minimal shell components. All observations above the top 
line were obtained during infall epochs.  
\textit{Panel 3}: The {\it dashed line} has the same meaning as in panel 1. 
Enhanced shell absorption is seen between phases 0.3 and 0.7.
\textit{Panel 4}: The velocity of the center-of-mass of the system 
is indicted by a {\it dashed line} and the radial velocity curve of the Be star
with a {\it solid line}. Only observations showing a 
shell component are included. Note the obvious presence of infalling material at
most phases except perhaps near phase 0.5 where material outflow is commonly seen 
in close binaries with early type primaries (Peters~2001). The errors for all 
observations are less than or equal to the dimension of the plotted point.    
\label{fig11}} 
\end{figure}

\begin{figure} 
\begin{center} 
{\includegraphics[height=16cm]{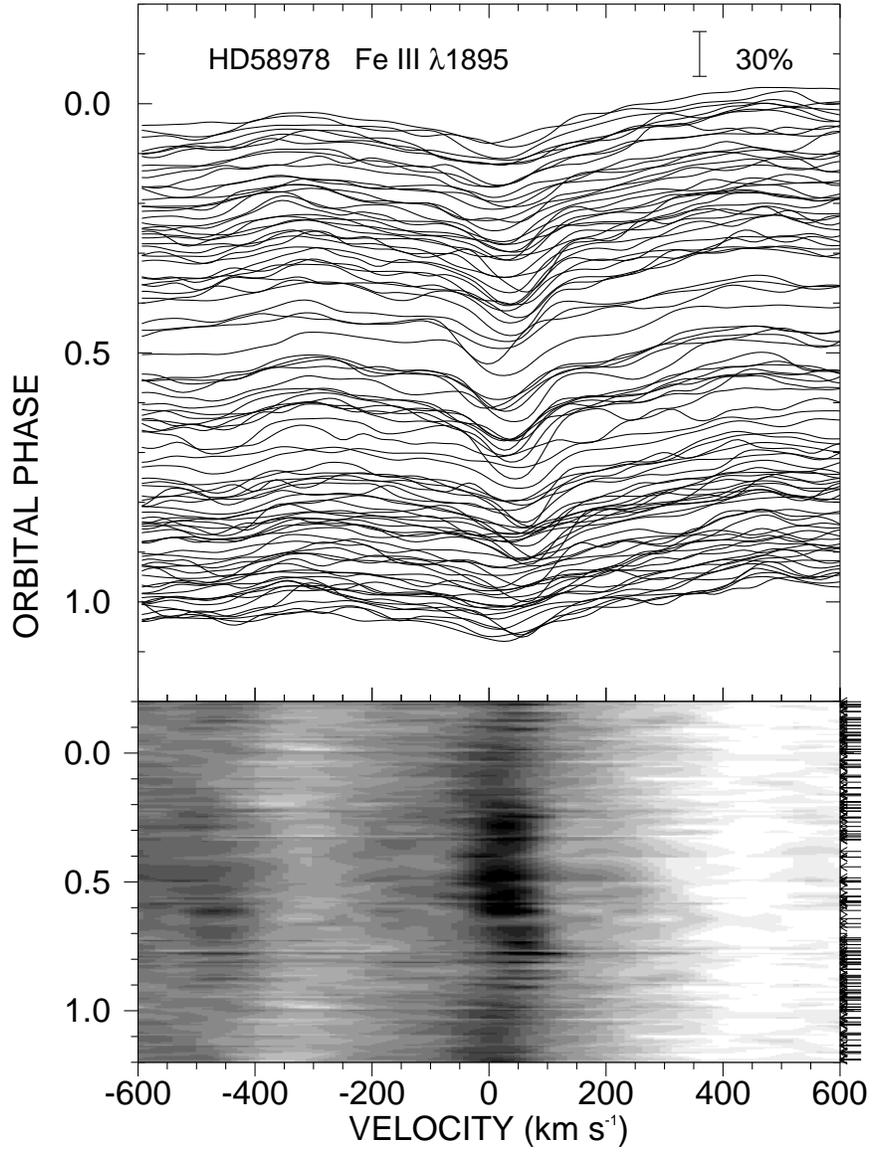}} 
\end{center} 
\caption{The orbital phase variations in \ion{Fe}{3} $\lambda 1895$ line  
plotted in the same format as Fig.~5.  A narrow ``shell'' feature can be seen  
around orbital phase 0.5. 
\label{fig12}} 
\end{figure}

\placefigure{fig10}     
\placefigure{fig11}     
\placefigure{fig12}     


Finally, we return to the cyclic variation in the \ion{N}{5} $\lambda1240$ 
wind line that first led \citet{riv04} to find the orbital period
of the binary.  The plot of these variations in Figure~13 shows that 
the greatest extent of blueshifted absorption occurs around $\phi=0.0$
when the subdwarf is in the foreground.  We note that such a strong \ion{N}{5} 
wind feature is quite unusual for Be stars \citep*{gra87}, and the orbital 
phasing suggests instead that the feature may form through absorption of 
the Be star's flux by the wind of the hot subdwarf when it is in the foreground. 
The profile would mainly be affected by blueshifted wind absorption since 
any redshifted absorption component would probably by canceled by redshifted emission
in the subdwarf's wind.  This kind of orbital wind variation is not seen in the 
other major wind line, the \ion{C}{4} $\lambda 1550$ doublet, and this may suggest  
that the subdwarf is N-enriched and C-depleted as expected for a star that has 
been stripped down to near the CNO-burning core.  One problem with this 
explanation is that the largest wind speeds observed, $\approx 400$ km~s$^{-1}$,
are much smaller than the wind terminal velocity of a few thousand km~s$^{-1}$
expected for a radiatively driven wind from the subdwarf \citep{lam99}. However, 
it may be that the wind optical depths are too small for absorption to occur at 
higher outflow velocities.  
 
\begin{figure} 
\begin{center} 
{\includegraphics[height=16cm]{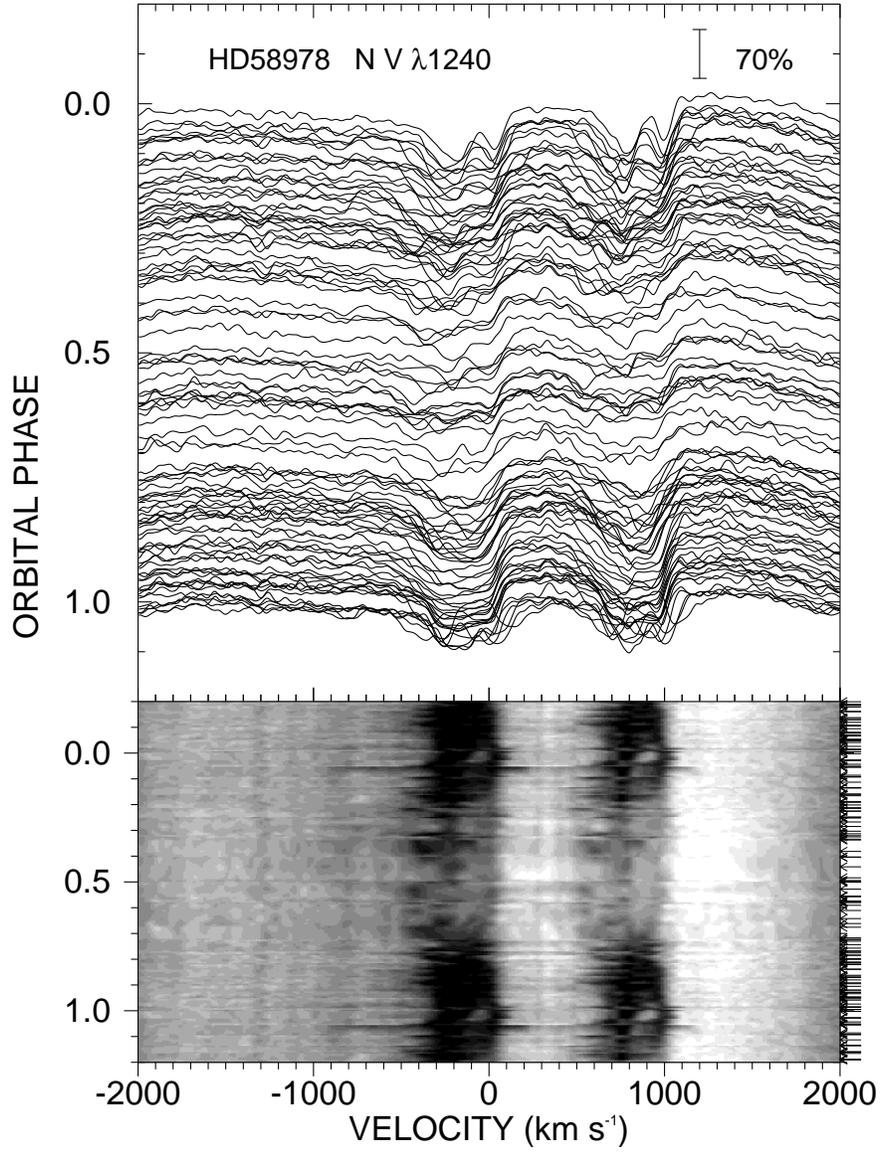}} 
\end{center} 
\caption{The orbital phase variations in the \ion{N}{5} $\lambda 1240$  
doublet plotted in the same format as Fig.~5 (in the velocity frame
of the blue member of the doublet).  The strong absorption occurs 
when the Be star is viewed through the wind of the hot subdwarf.  
\label{fig13}} 
\end{figure}

\placefigure{fig13}     
 
In 1995 March a sequence of daily {\it IUE} SWP HIRES spectra of FY~CMa 
were obtained over the course of 16 days.  Fortuitously these spectra 
were centered in time around orbital phase 0.0 and extended from  
phases 0.784 to 0.213.  The temporal variations in the \ion{N}{5} $\lambda1240$ 
doublet are illustrated in a gray-scale diagram in Figure~14. More detail can be seen
from the plot of selected observations shown in Figure~15.  The \ion{N}{5} 
features initially showed very little structure until a core with the velocity of
the center-of-mass of the system appeared around phase 0.97.  The core, still
with no velocity shift, reached its greatest strength at conjunction. 
It broadened by phase 0.08, and had completely vanished by phase 0.21.
The core is probably formed in a shock-heated region between the stars 
where the N-enhanced wind from the subdwarf collides with the disk of 
the Be star.

\begin{figure} 
\begin{center} 
{\includegraphics[height=16cm]{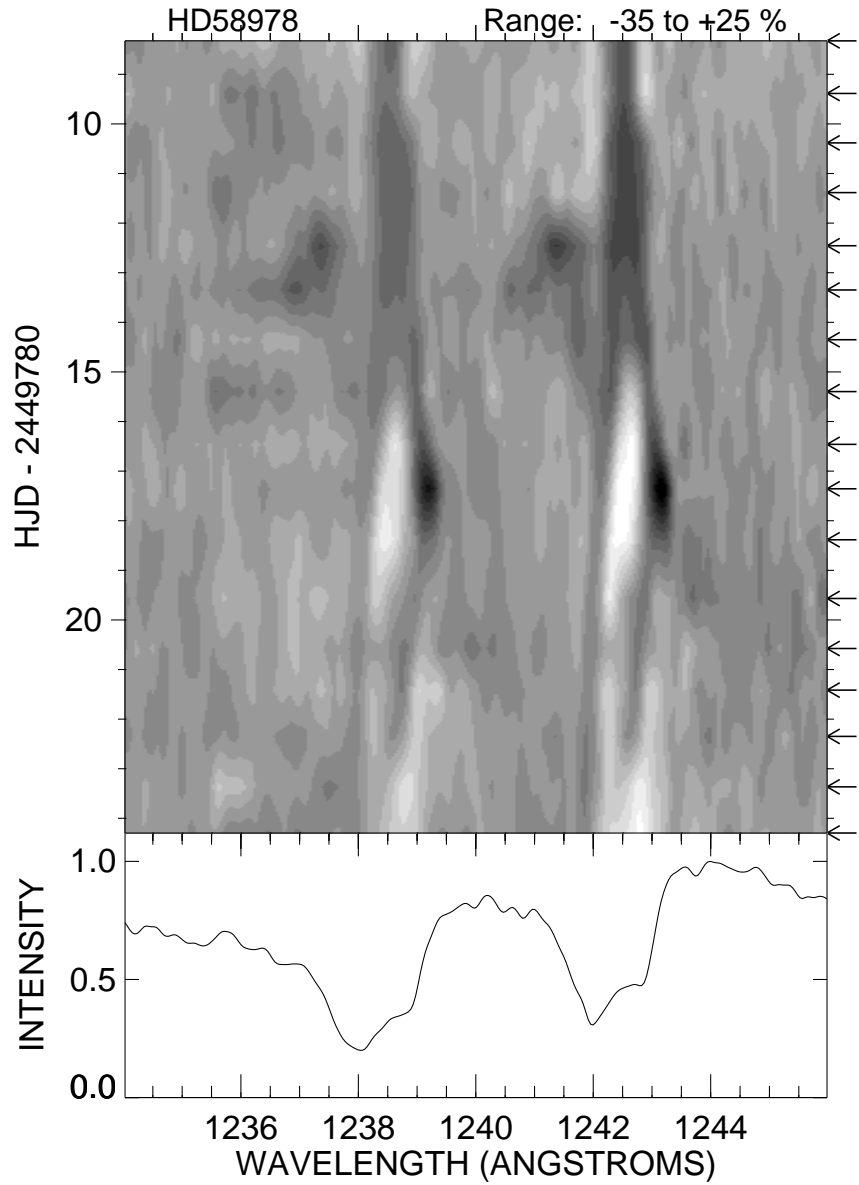}} 
\end{center} 
\caption{The temporal variations in the \ion{N}{5} $\lambda 1240$  
doublet observed in a daily sequence of {\it IUE} observations obtained 
in 1995 March. The lower portion shows the average profile while the upper
panel shows a gray-scale representation of the deviations from
the average as a function of time.  A shell feature 
appeared near orbital phase 0.0 at HJD~2449796.4. 
\label{fig14}} 
\end{figure}

\begin{figure} 
\begin{center} 
{\includegraphics[height=16cm]{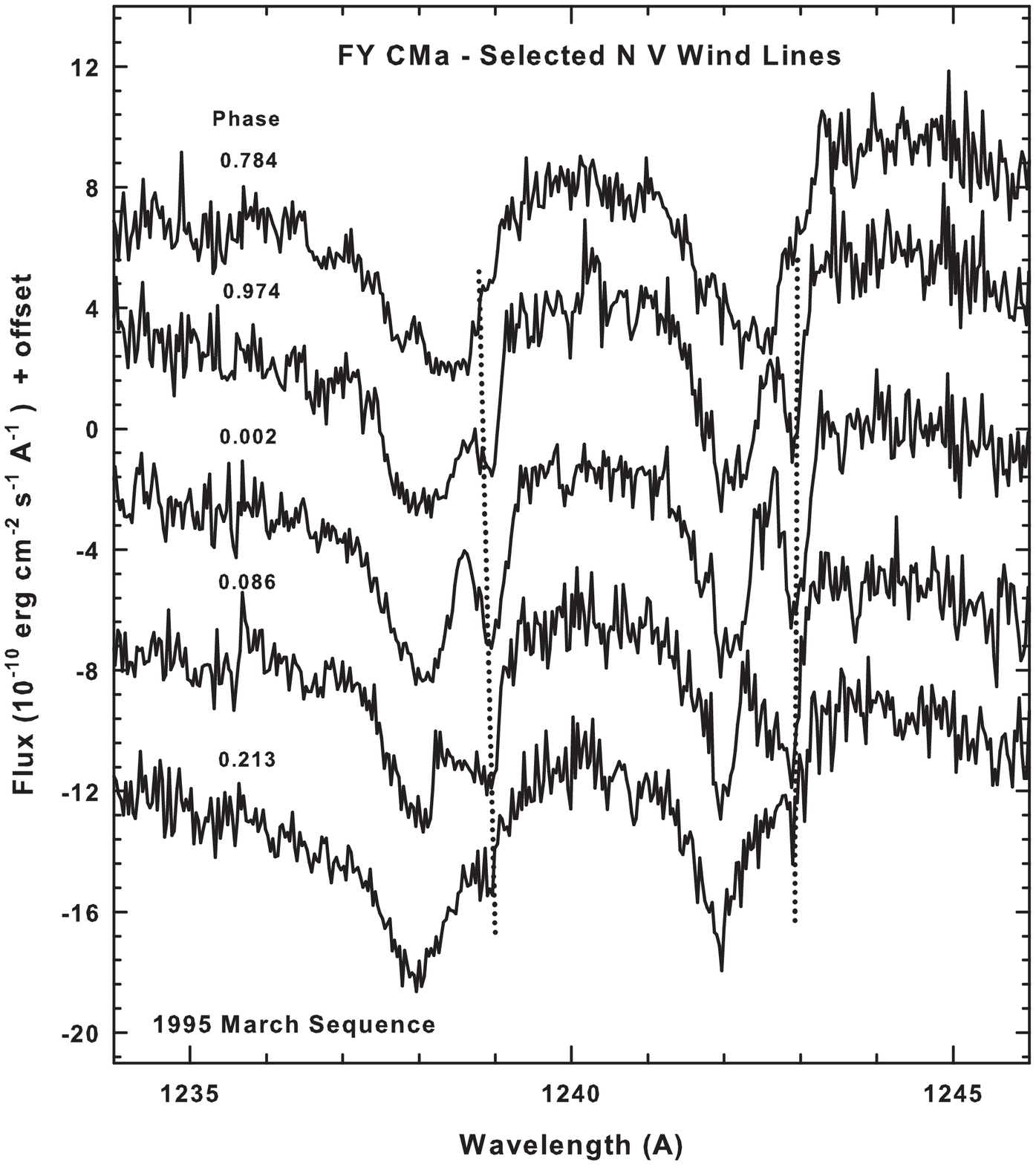}} 
\end{center} 
\caption{Selected profiles of the \ion{N}{5} doublet observed during the 1995 March
sequence show the details of the development and fading of the \ion{N}{5} shell 
component. The feature was strongest at inferior conjunction of the subdwarf and 
essentially absent near the quadrature points. The vertical {\it dotted} lines show 
the locations of the observed line centers in the velocity frame of the center-of-mass 
of the system. The y-axis flux offsets are (from top to bottom) $0, -4, -10, -15,$ and 
$-20$ in the flux units plotted.
\label{fig15}} 
\end{figure}

\placefigure{fig14}     
\placefigure{fig15}     

Our work demonstrates that FY~CMa represents the third known case of 
a Be star with a hot subdwarf companion, the remains of an intense 
binary interaction that stripped down the mass donor to a tiny fraction 
of its original mass and left the mass gainer spinning rapidly. 
The subdwarf secondary in FY~CMa has a mass that is close to the 
Chandrasekhar limit, and lacking a significant H-envelope, it may represent 
the progenitor of a supernova type SN~Ibc.  The system is now wide enough 
that it is unlikely that the Be star will be substantially spun down by tidal 
effects, so its rapid rotation will be important throughout its subsequent evolution. 
Such rapidly rotating massive stars may be related to the progenitors of   
collapsars and gamma-ray bursters \citep{can07}.  
 
 
\acknowledgments 
 
We thank Daryl Willmarth and the staff of KPNO for their assistance  
in making these observations possible and  
Diana Gudkova for her assistance in organizing the optical spectra.
The {\it IUE} data presented in this paper were obtained from  
the Multimission Archive at the Space Telescope Science  
Institute (MAST). STScI is operated by the Association of  
Universities for Research in Astronomy, Inc., under NASA  
contract NAS5-26555. Support for MAST for non-HST data is  
provided by the NASA Office of Space Science via grant  
NAG5-7584 and by other grants and contracts. 
Many researchers contributed to the 97 {\it IUE} SWP HIRES images
used in this study, including programs by K. S. Bjorkman,
C. A. Grady, H. Henrichs,  J. Nichols, G. J. Peters, and C.-C. Wu. 
We especially thank Carol Grady, who had the foresight to obtain the 
unique set of daily {\it IUE} spectra in 1995, more than a decade
before analysis software allowed us to suggest an interpretation for the
striking short-term variability seen in the \ion{N}{5} doublet and other
wind features. We also thank an anonymous referee whose suggestions led to 
several improvements in the paper. This work was supported in part by the 
National Science Foundation under grant AST-0606861. 
Institutional support has been provided from the USC 
Women in Science \& Engineering (WiSE) program and from the GSU College 
of Arts and Sciences and the Research Program Enhancement 
fund of the Board of Regents of the University System of Georgia, 
administered through the GSU Office of the Vice President 
for Research.  

{\it Facilities:} \facility{IUE }, \facility{KPNO:CFT } 
 
 

\clearpage


 
\begin{deluxetable}{ccccccl} 
\tabletypesize{\scriptsize} 
\tablewidth{0pc} 
\tablecaption{Journal of Spectroscopy \label{tab1}} 
\tablehead{ 
\colhead{Dates} & 
\colhead{Besselian} & 
\colhead{Range} & 
\colhead{Resolving Power} & 
\colhead{Number of} & 
\colhead{Telescope/} \\ 
\colhead{(HJD-2,400,000)} & 
\colhead{Year} & 
\colhead{(\AA)} & 
\colhead{($\lambda/\triangle\lambda$)} & 
\colhead{Spectra} & 
\colhead{Grating/Detector}} 
\startdata 
44170 -- 49804   &  1979.8 -- 1995.2 &   1200 -- 1900    & 10000    &   97 & {\it IUE}/Echelle/SWP  \\
46123            &  1985.2           &   6477 -- 6654    & 13000    &\phn1 &  KPNO CF/B/TI3  \\
46153            &  1985.2           &   6514 -- 6691    & 13000    &\phn1 &  KPNO CF/B/TI3  \\
46735            &  1986.8           &   6511 -- 6688    & 13000    &\phn1 &  KPNO CF/B/TI3  \\
46903 -- 46906   &  1987.3 -- 1987.3 &   6522 -- 6698    & 13000    &\phn3 &  KPNO CF/B/TI3  \\
46918 -- 46922   &  1987.3 -- 1987.3 &   6520 -- 6698    & 13000    &\phn6 &  KPNO CF/B/TI3  \\
47033 -- 47034   &  1987.6 -- 1987.7 &   6518 -- 6695    & 13000    &\phn2 &  KPNO CF/B/TI3  \\
47303 -- 47307   &  1988.4 -- 1988.4 &   6517 -- 6694    & 13000    &\phn9 &  KPNO CF/B/TI3  \\
47470 -- 47473   &  1988.8 -- 1988.9 &   6495 -- 6717    & 11000    &   11 &  KPNO CF/B/RCA2  \\
47561 -- 47562   &  1989.1 -- 1989.1 &   6514 -- 6690    & 14000    &\phn3 &  KPNO CF/B/TI3  \\
47637 -- 47641   &  1989.3 -- 1989.3 &   6520 -- 6697    & 14000    &\phn4 &  KPNO CF/B/TI3  \\
47940 -- 47941   &  1990.1 -- 1990.1 &   6523 -- 6699    & 14000    &\phn2 &  KPNO CF/B/TI3  \\
47982 -- 47984   &  1990.2 -- 1990.3 &   6520 -- 6697    & 14000    &\phn2 &  KPNO CF/B/TI3  \\
48314 -- 48315   &  1991.2 -- 1991.2 &   6542 -- 6690    & 18000    &\phn3 &  KPNO CF/A/ST1K \\
48319 -- 48322   &  1991.2 -- 1991.2 &   6520 -- 6696    & 14000    &\phn4 &  KPNO CF/B/TI3  \\
48515 -- 48517   &  1991.7 -- 1991.7 &   6517 -- 6693    & 14000    &\phn5 &  KPNO CF/B/TI3  \\
49058 -- 49061   &  1993.2 -- 1993.2 &   6384 -- 6723    & 19000    &\phn3 &  KPNO CF/A/T2KB \\
49444 -- 49448   &  1994.2 -- 1994.3 &   6525 -- 6693    & 18000    &\phn4 &  KPNO CF/A/T1KA \\
51123 -- 51126   &  1998.8 -- 1998.9 &   6359 -- 6677    & 20000    &\phn4 &  KPNO CF/A/F3KB \\
51192 -- 51197   &  1999.0 -- 1999.0 &   6420 -- 6737    & 20000    &\phn5 &  KPNO CF/A/F3KB \\
51502 -- 51512   &  1999.9 -- 1999.9 &   6415 -- 6733    & 20000    &   16 &  KPNO CF/A/F3KB \\
51613 -- 51614   &  2000.2 -- 2000.2 &   6459 -- 6776    & 20000    &\phn8 &  KPNO CF/A/F3KB \\
51615 -- 51618   &  2000.2 -- 2000.2 &   6413 -- 6730    & 20000    &\phn9 &  KPNO CF/A/F3KB \\
51818 -- 51831   &  2000.7 -- 2000.8 &   6440 -- 7105    &\phn9500  &\phn9 &  KPNO CF/B/F3KB \\
51850 -- 51852   &  2000.8 -- 2000.8 &   6413 -- 6733    & 20000    &\phn7 &  KPNO CF/A/F3KB \\
51890 -- 51902   &  2000.9 -- 2001.0 &   6440 -- 7105    &\phn9500  &   12 &  KPNO CF/B/F3KB \\
51914 -- 51918   &  2001.0 -- 2001.0 &   6432 -- 6749    & 20000    &   18 &  KPNO CF/A/F3KB \\
53292 -- 53295   &  2004.8 -- 2004.8 &   6470 -- 7140    &\phn9500  &\phn4 &  KPNO CF/B/T2KB \\
54021 -- 54025   &  2006.8 -- 2006.8 &   6473 -- 7143    &\phn9500  &\phn2 &  KPNO CF/B/T2KB \\
\enddata 
\end{deluxetable} 
\newpage

\begin{deluxetable}{lccccccc} 
\tabletypesize{\scriptsize} 
\tablewidth{0pc} 
\tablenum{2} 
\tablecaption{{\it IUE} Radial Velocity Measurements\label{tab2}} 
\tablehead{ 
\colhead{Date}             & 
\colhead{UT}               & 
\colhead{SWP}              & 
\colhead{Orbital}          & 
\colhead{$V_1$}            & 
\colhead{$(O-C)_1$}        & 
\colhead{$V_2$}            & 
\colhead{$(O-C)_2$}        \\  
\colhead{(HJD--2,400,000)} & 
\colhead{(yyyy-mm-dd)}    & 
\colhead{Number}           & 
\colhead{Phase}            & 
\colhead{(km s$^{-1}$)}    & 
\colhead{(km s$^{-1}$)}    & 
\colhead{(km s$^{-1}$)}    & 
\colhead{(km s$^{-1}$)}    } 
\startdata 
 44170.044 \dotfill & 1979-10-23 &  \phn6963 &  0.985 &  
\phn\phs     $  51.5$ &\phs     $  16.6$ & 
\phn\phn\phs $   1.6$ &         $ -17.9$ \\ 
 44330.226 \dotfill & 1980-03-31 &  \phn8617 &  0.285 &  
\phn\phs     $  23.9$ &\phn\phs $   4.4$ & 
\phs         $ 159.8$ &\phn\phs $   3.4$ \\ 
 44853.156 \dotfill & 1981-09-05 & 14910 &  0.321 &  
\phn\phs     $  22.2$ &\phn\phs $   1.6$ & 
\phs         $ 138.6$ &\phn     $  -8.4$ \\ 
 44867.935 \dotfill & 1981-09-20 & 15053 &  0.717 &  
\phn\phs     $  40.5$ &\phn     $  -7.3$ & 
             $-123.5$ &         $ -29.2$ \\ 
 44920.747 \dotfill & 1981-11-12 & 15478 &  0.135 &  
\phn\phs     $  39.0$ &\phs     $  16.2$ & 
             \nodata  &         \nodata  \\ 
\enddata 
\tablecomments{A machine readable version of the full Table 2 is available 
in the electronic edition of the {\it Astrophysical Journal.}  A portion 
is shown here for guidance regarding its form and content.}
\end{deluxetable}
 
\newpage

\begin{deluxetable}{lcc} 
\tablewidth{0pc} 
\tablenum{3} 
\tablecaption{Orbital Elements for FY CMa\label{tab3}} 
\tablehead{ 
\colhead{Element} & 
\colhead{H$\alpha$ Wings} & 
\colhead{{\it IUE} ccfs}} 
\startdata 
$P$~(days)                \dotfill & $37.253 \pm 0.007$ & $37.257 \pm 0.003$   \\ 
$T_{SC}$ (HJD--2,448,000) \dotfill & $530.19 \pm 0.76$  & $529.64 \pm 0.15$    \\ 
$K_1$ (km s$^{-1}$)       \dotfill & $17.4 \pm  0.9$    & $14.4 \pm  0.9$      \\ 
$K_2$ (km s$^{-1}$)       \dotfill & \nodata            & $128.2 \pm 2.2$      \\ 
$\gamma_1$ (km s$^{-1}$)  \dotfill & $47.1 \pm  0.6$    & $33.6 \pm 0.7$       \\ 
$\gamma_2$ (km s$^{-1}$)  \dotfill & \nodata            & $31.2 \pm 1.7$       \\ 
$M_2/M_1$                 \dotfill & \nodata            & $0.113 \pm 0.007$    \\ 
$M_1\sin ^3 i$ ($M_\odot$)\dotfill & \nodata            & $10.1 \pm 0.4$       \\ 
$M_2\sin ^3 i$ ($M_\odot$)\dotfill & \nodata            & $1.13 \pm 0.08$      \\ 
$a\sin i$ ($R_\odot$)     \dotfill & \nodata            & $105.0 \pm 1.7$      \\ 
rms$_1$ (km s$^{-1}$)     \dotfill & 7.6                & 6.5                  \\ 
rms$_2$ (km s$^{-1}$)     \dotfill & \nodata            & 14.5                 \\ 
\enddata 
\end{deluxetable} 
\newpage 
 
\begin{deluxetable}{lcc} 
\tablewidth{0pc} 
\tablenum{4} 
\tablecaption{Stellar Parameters for FY CMa\label{tab4}} 
\tablehead{ 
\colhead{Parameter} & 
\colhead{Primary} & 
\colhead{Secondary} } 
\startdata 
$T_{\rm eff}$ (kK)        & $27.5 \pm 3.0$ & $45 \pm 5$ \\ 
$\log g$ (cgs)            & $4.0$          & $4.3 \pm 0.6$ \\ 
$V\sin i$ (km s$^{-1}$)   & $340 \pm 40$   & $41 \pm 5$ \\ 
$M/M_\odot$	          & $10 - 13$      & $1.1 - 1.5$ \\ 
$R/R_\odot$	          & $4.1 - 7.6$    & $0.5 - 0.9$ \\ 
\enddata 
\end{deluxetable} 
\newpage

\begin{deluxetable}{lccccc} 
\tabletypesize{\scriptsize} 
\tablewidth{0pc} 
\tablenum{5} 
\tablecaption{Red Spectra Measurements\label{tab5}} 
\tablehead{ 
\colhead{Date}             & 
\colhead{Orbital}          & 
\colhead{$W_\lambda$(H$\alpha$)} & 
\colhead{$V$(H$\alpha$ wings)}   & 
\colhead{$V$(He I peaks)} & 
\colhead{$\triangle V$(He I peaks)}        \\  
\colhead{(HJD--2,400,000)} & 
\colhead{Phase}            & 
\colhead{(\AA )}    & 
\colhead{(km s$^{-1}$)}    & 
\colhead{(km s$^{-1}$)}    & 
\colhead{(km s$^{-1}$)}    } 
\startdata 
 46122.805 \dotfill &  0.399 &  
         $-10.98$ &\phn\phs     $  26.8$ &          \nodata    &        \nodata \\ 
 46152.702 \dotfill &  0.201 &  
         $-13.32$ &\phn\phs     $  21.8$ &\phs         $  216$ &        \nodata \\ 
 46734.931 \dotfill &  0.829 &  
         $-11.22$ &\phn\phs     $  71.1$ &             $ -154$ &        \nodata \\ 
 46902.728 \dotfill &  0.333 &  
         $-11.76$ &\phn\phs     $  30.2$ &          \nodata    &        \nodata \\ 
 46903.703 \dotfill &  0.359 &  
         $-12.39$ &\phn\phs     $  25.2$ &          \nodata    &        \nodata \\ 
\enddata 
\tablecomments{Table 5 is available in its entirely in the electronic edition of the 
{\it Astrophysical Journal.}  A portion is shown here for guidance regarding its 
form and content.}
\end{deluxetable} 


\clearpage



\clearpage

{\Huge 

\begin{center}
ONLINE MATERIAL
\end{center}

}

\clearpage

\begin{verbatim}

Title: Detection of a Hot Subdwarf Companion to the Be Star FY CMa
Authors: G. J. Peters, D. R. Gies, E. D. Grundstrom, M. V. McSwain
Table: IUE Radial Velocity Measurements
================================================================================
Byte-by-byte Description of file: tab2.txt
--------------------------------------------------------------------------------
   Bytes Format Units Label  Explanations
--------------------------------------------------------------------------------
   1-  9 F9.3   d       Date   HJD-2400000
  11- 14 I4     --      Obs.y  Observation UT year
  16- 17 I2     --      Obs.m  Observation UT month
  19- 20 I2     --      Obs.d  Observation UT day
  21- 26 I6     --      SWP    IUE SWP number
  27- 32 F6.3   --      Phase  Orbital phase from Be star superior conjunction
  33- 38 F6.1   km/s    HRV1   Heliocentric radial velocity for Be star
  39- 44 F6.1   km/s    OMC1   Observed minus calculated velocity for Be star
  45- 51 A7     km/s    HRV2   Heliocentric radial velocity for subdwarf
  52- 57 A6     km/s    OMC2   Observed minus calculated velocity for subdwarf
--------------------------------------------------------------------------------
44170.044 1979-10-23  6963 0.985  51.5  16.6    1.6 -17.9
44330.226 1980-03-31  8617 0.285  23.9   4.4  159.8   3.4
44853.156 1981-09-05 14910 0.321  22.2   1.6  138.6  -8.4
44867.935 1981-09-20 15053 0.717  40.5  -7.3 -123.5 -29.2
44920.747 1981-11-12 15478 0.135  39.0  16.2    ?     ?  
44972.649 1982-01-03 15933 0.528  36.8   0.7    ?     ?  
44974.604 1982-01-05 15957 0.580  33.9  -6.7    ?     ?  
44975.736 1982-01-06 15979 0.611  34.4  -8.5  -67.0 -16.0
45571.096 1983-08-24 20769 0.591  43.7   2.3  -48.6 -10.7
45573.015 1983-08-26 20805 0.642  52.5   7.6    ?     ?  
45576.057 1983-08-29 20837 0.724  57.4   9.5 -105.8 -10.5
45578.924 1983-09-01 20874 0.801  50.7   3.4    ?     ?  
46338.986 1985-09-30 26810 0.201  23.1   3.3  146.0  -7.5
46340.871 1985-10-02 26829 0.252   8.4 -10.8    ?     ?  
46342.778 1985-10-04 26847 0.303  15.0  -5.0    ?     ?  
46343.927 1985-10-05 26868 0.334  11.1 -10.0  142.2   0.2
46430.639 1985-12-31 27426 0.661  52.3   6.4  -74.0   3.7
46436.586 1986-01-06 27459 0.821  37.1  -9.5  -61.0  23.4
46440.366 1986-01-09 27479 0.923  40.8   0.4  -24.4   4.4
46440.609 1986-01-10 27484 0.929  34.0  -5.8  -16.1   8.0
46440.784 1986-01-10 27488 0.934  29.2 -10.3  -21.3  -0.7
46441.374 1986-01-10 27496 0.950  38.5   0.4    ?     ?  
46441.513 1986-01-11 27499 0.953  35.5  -2.3    ?     ?  
46441.720 1986-01-11 27504 0.959  36.6  -0.7   10.0  11.6
46590.362 1986-06-08 28457 0.949  41.6   3.4   -8.5   1.1
46691.971 1986-09-18 29238 0.676  49.1   2.6    ?     ?  
46705.946 1986-10-02 29349 0.051  27.2  -1.9    ?     ?  
46709.960 1986-10-06 29386 0.159  25.0   3.5    ?     ?  
46744.776 1986-11-10 29646 0.093  29.0   3.4    ?     ?  
46823.765 1987-01-28 30182 0.213  19.3  -0.2  145.4 -10.6
46823.786 1987-01-28 30183 0.214  20.7   1.2    ?     ?  
46852.535 1987-02-26 30392 0.985  40.3   5.4   15.9  -3.6
46861.455 1987-03-06 30443 0.225  11.5  -7.8  163.1   5.3
46919.233 1987-05-03 30901 0.776  35.9 -12.0    ?     ?  
47060.818 1987-09-22 31900 0.576  31.4  -8.8    ?     ?  
47085.860 1987-10-17 32114 0.248  16.8  -2.4    ?     ?  
47103.822 1987-11-04 32222 0.730  45.6  -2.3  -81.4  14.6
47106.832 1987-11-07 32265 0.811  47.1   0.1    ?     ?  
47113.690 1987-11-14 32317 0.995  34.4   0.3   18.8  -8.4
47125.983 1987-11-26 32398 0.325  21.9   1.2    ?     ?  
47141.568 1987-12-12 32505 0.743  48.7   0.7    ?     ?  
47167.580 1988-01-07 32677 0.441  27.5  -0.9    ?     ?  
47169.553 1988-01-09 32688 0.494  28.6  -4.5    ?     ?  
47172.522 1988-01-12 32702 0.574  46.5   6.4  -25.6   0.7
47198.348 1988-02-06 32872 0.267  18.1  -1.1    ?     ?  
47206.425 1988-02-14 32925 0.484  30.4  -1.8    ?     ?  
47215.342 1988-02-23 32968 0.723  47.0  -0.9 -103.4  -8.2
47225.456 1988-03-04 33039 0.995  39.4   5.3   29.1   2.0
47240.674 1988-03-20 33119 0.403  27.7   2.3   92.9 -11.5
47258.449 1988-04-06 33221 0.880  42.9  -0.6  -69.2 -12.9
47262.396 1988-04-10 33245 0.986  29.1  -5.7   13.1  -7.2
47271.315 1988-04-19 33315 0.226  23.6   4.3  170.4  12.5
47293.289 1988-05-11 33508 0.816  47.9   1.0  -72.4  13.9
47305.145 1988-05-23 33616 0.134  17.1  -5.7    ?     ?  
47307.279 1988-05-25 33640 0.191  10.7  -9.4  171.8  21.1
47426.980 1988-09-22 34288 0.404  29.6   4.2    ?     ?  
47460.861 1988-10-26 34606 0.313  19.8  -0.5    ?     ?  
47478.914 1988-11-13 34736 0.798  48.7   1.3    ?     ?  
47514.647 1988-12-19 35074 0.757  38.1  -9.9 -110.3 -13.4
47515.683 1988-12-20 35082 0.785  55.1   7.4    ?     ?  
47517.675 1988-12-22 35094 0.838  54.9   9.0    ?     ?  
47518.576 1988-12-23 35101 0.862  36.4  -8.2    ?     ?  
47519.662 1988-12-24 35112 0.892  38.4  -4.3    ?     ?  
47537.921 1989-01-11 35320 0.382  18.1  -5.7  108.1  -9.9
47581.733 1989-02-24 35614 0.558  35.8  -2.9   27.8  42.0
47619.671 1989-04-03 35919 0.576  44.3   4.1  -17.7   9.9
47656.154 1989-05-09 36229 0.555  40.2   1.7    ?     ?  
47908.837 1990-01-17 38034 0.337  18.9  -2.4    ?     ?  
47928.730 1990-02-06 38148 0.871  44.4   0.3    ?     ?  
47999.467 1990-04-17 38626 0.770  48.7   0.8 -103.3  -7.3
48004.512 1990-04-23 38655 0.905  43.1   1.4    ?     ?  
48175.728 1990-10-11 39806 0.501  21.4 -12.3   42.8  12.3
48232.662 1990-12-07 40284 0.029  36.1   5.1   36.2 -18.3
48297.434 1991-02-09 40846 0.768  42.1  -5.9    ?     ?  
48349.522 1991-04-03 41276 0.166  10.4 -10.7    ?     ?  
49313.649 1993-11-22 49290 0.043  29.7  -0.0   67.6   1.8
49442.353 1994-03-30 50426 0.498  31.3  -2.1    ?     ?  
49470.349 1994-04-27 50635 0.249  22.7   3.5    ?     ?  
49686.641 1994-11-30 52938 0.055  29.0   0.3    ?     ?  
49762.511 1995-02-14 53907 0.091  29.1   3.3   90.1 -10.7
49788.326 1995-03-11 54102 0.784  56.3   8.6 -100.7  -6.6
49789.386 1995-03-12 54111 0.813  54.9   7.9  -93.4  -6.2
49790.383 1995-03-13 54123 0.839  52.7   6.9  -92.8 -15.5
49791.385 1995-03-14 54134 0.866  45.9   1.5    ?     ?  
49792.457 1995-03-15 54147 0.895  39.0  -3.5    ?     ?  
49793.349 1995-03-16 54155 0.919  51.6  10.9    2.7  34.0
49794.351 1995-03-17 54163 0.946  46.0   7.6   26.9  38.4
49795.397 1995-03-18 54173 0.974  18.6 -17.4    ?     ?  
49796.460 1995-03-19 54183 0.002  31.7  -1.7   32.8  -0.4
49797.356 1995-03-20 54187 0.027  20.9 -10.3   49.8  -2.7
49798.385 1995-03-21 54202 0.054  32.2   3.4   72.6  -1.4
49799.568 1995-03-23 54210 0.086  35.8   9.6   87.2  -9.9
49800.577 1995-03-24 54215 0.113  25.6   1.4    ?     ?  
49801.415 1995-03-24 54220 0.135  35.6  12.9    ?     ?  
49802.348 1995-03-25 54223 0.161  31.7  10.3  142.7   3.0
49803.365 1995-03-26 54230 0.188  20.1  -0.2    ?     ?  
49804.296 1995-03-27 54238 0.213  23.5   3.9  154.1  -1.8

\end{verbatim}

\clearpage

\begin{verbatim}

Title: Detection of a Hot Subdwarf Companion to the Be Star FY CMa
Authors: G. J. Peters, D. R. Gies, E. D. Grundstrom, M. V. McSwain
Table: Red Spectra Measurements
================================================================================
Byte-by-byte Description of file: tab5.txt
--------------------------------------------------------------------------------
   Bytes Format Units Label      Explanations
--------------------------------------------------------------------------------
   1-  9 F9.3   d       Date     HJD-2400000
  10- 15 F6.3   --      Phase    Orbital phase from Be star superior conjunction
  16- 22 F7.2   0.1nm   EQW      H-alpha equivalent width (Angstroms)
  23- 28 F6.1   km/s    HRV(HA)  Heliocentric radial velocity H-alpha wings
  29- 33 A5     km/s    HRV(HE)  Heliocentric radial velocity He I peaks
  34- 37 A4     km/s    DVR(HE)  He I peak velocity separation
--------------------------------------------------------------------------------
Note (1): HRV(HA) values omitted in Fig. 1 and in the orbital solution of 
Table 3, column 2, for HJD-2400000 = 47639.671 and 47983.611.
--------------------------------------------------------------------------------
46122.805 0.399 -10.98  26.8   ?   ? 
46152.702 0.201 -13.32  21.8  216  ? 
46734.931 0.829 -11.22  71.1 -154  ? 
46902.728 0.333 -11.76  30.2   ?   ? 
46903.703 0.359 -12.39  25.2   ?   ? 
46905.682 0.412 -13.46  34.7   ?   ? 
46917.654 0.733 -13.94  56.7  -81  56
46918.676 0.761 -14.15  42.9  -88  81
46919.674 0.787 -13.37  56.3  -79  94
46920.675 0.814 -11.87  52.9  -75 102
46921.670 0.841 -11.76  46.4  -70 115
46921.684 0.841 -12.86  43.3  -70 120
47033.012 0.830  -8.52  62.2  -68 107
47034.004 0.856 -12.34  53.6  -63 113
47302.650 0.067 -15.87  33.1   ?   ? 
47303.636 0.093 -15.65  32.5  143  ? 
47303.642 0.093 -14.93  40.4  147  ? 
47304.622 0.120 -13.67  35.4  160  ? 
47304.631 0.120 -14.61  39.6  158  ? 
47305.623 0.147 -11.51  45.5  171  ? 
47305.633 0.147 -13.28  42.0  171  ? 
47306.622 0.173 -10.57  44.4  173  ? 
47306.632 0.174 -12.96  41.2  175  ? 
47469.823 0.554 -13.56  49.0  -94  ? 
47469.831 0.554 -13.43  47.0  -99  ? 
47469.914 0.556 -13.15  52.8  -98  ? 
47469.997 0.559 -13.47  54.1  -92  ? 
47470.026 0.559 -13.63  54.6  -96  ? 
47470.826 0.581 -11.53  57.0 -100  ? 
47471.014 0.586 -11.66  54.7 -107  ? 
47471.846 0.608 -11.87  57.9 -111  ? 
47471.995 0.612 -12.00  62.8 -124  ? 
47472.848 0.635 -11.39  66.6 -128  ? 
47473.033 0.640 -11.66  66.3 -132  ? 
47560.654 0.992 -11.66  49.9   ?   ? 
47561.681 0.019 -15.26  46.5   ?   ? 
47561.792 0.022 -14.90  46.4   ?   ? 
47636.683 0.033 -10.94  36.0   ?   ? 
47639.671 0.113 -10.92 -13.9  203  ? 
47639.680 0.113 -11.53  33.1  207  ? 
47640.635 0.139 -12.13  32.1  207  ? 
47939.622 0.164 -10.96  18.8  226  ? 
47940.860 0.197 -11.92  10.1  233  ? 
47981.708 0.293  -5.32  10.3   ?   ? 
47983.611 0.344 -10.74   4.4   ?   ? 
48313.702 0.204 -13.26  28.5   ?   ? 
48313.788 0.207 -13.20  29.4  175  ? 
48314.619 0.229 -12.31  35.4  173  ? 
48318.615 0.336 -12.59  20.8  138 113
48319.800 0.368 -15.66  26.3  122 124
48320.616 0.390 -12.22  26.3  113 132
48321.610 0.416 -12.08  43.9  113 137
48514.969 0.606 -12.09  50.5  -66  ? 
48515.030 0.608 -10.42  53.9  -68  ? 
48515.972 0.633 -11.30  58.2  -96  ? 
48516.017 0.634 -11.93  53.9  -92  ? 
48517.021 0.661 -12.21  58.6   ?   ? 
49058.749 0.202 -14.58  34.4  179  ? 
49061.769 0.283 -14.06  39.0  162  94
49061.773 0.283 -15.28  30.5  160  94
49443.771 0.536 -14.61  42.9  -52  ? 
49444.729 0.562 -10.59  49.0  -68  ? 
49445.749 0.589 -10.22  52.3  -77  ? 
49447.755 0.643 -13.23  51.7  -96  ? 
51123.027 0.608 -16.17  52.1   ?   ? 
51123.034 0.609 -15.94  53.7   ?   ? 
51124.020 0.635 -16.40  56.9   ?   ? 
51126.033 0.689 -16.83  55.3   ?   ? 
51191.897 0.457 -13.75  41.1   ?   ? 
51192.937 0.485 -14.95  41.7   ?   ? 
51193.927 0.512 -13.53  46.0   ?   ? 
51196.937 0.592 -13.35  53.9  -68  ? 
51197.933 0.619 -14.21  55.1  -73  ? 
51502.936 0.806 -13.04  59.8 -141  ? 
51503.977 0.833 -14.12  59.4 -137  ? 
51504.994 0.861 -13.71  57.7 -112  ? 
51506.025 0.888 -13.94  54.6 -132  ? 
51507.019 0.915 -14.53  49.8 -130  ? 
51508.023 0.942 -14.07  47.1  -53 124
51509.013 0.969 -13.77  46.8  -14 100
51510.006 0.995 -15.13  46.0   -6 120
51510.053 0.997 -15.52  44.6   -2 111
51510.798 0.017 -15.91  43.7   72  ? 
51510.847 0.018 -15.76  42.2   73  ? 
51510.918 0.020 -15.41  44.6   78  ? 
51510.969 0.021 -15.33  43.5   83  ? 
51511.946 0.047 -16.08  40.7   97  ? 
51512.015 0.049 -15.34  41.3   98  ? 
51512.037 0.050 -15.39  40.7  100  ? 
51612.632 0.750 -12.91  69.1 -162  ? 
51612.693 0.751 -13.83  69.8 -160  ? 
51612.802 0.754 -12.84  69.0 -159  ? 
51612.822 0.755 -11.87  66.8 -154  ? 
51613.617 0.776 -12.95  66.5 -157  ? 
51613.703 0.779 -12.81  67.5 -152  ? 
51613.712 0.779 -13.04  67.3 -160  ? 
51613.751 0.780 -13.07  69.2 -158  ? 
51614.612 0.803 -13.18  69.5 -156  ? 
51614.829 0.809 -13.68  69.7 -160  ? 
51615.607 0.830 -13.24  68.1 -158  ? 
51615.717 0.833 -13.08  67.0 -160  ? 
51615.759 0.834 -14.07  67.1 -160  ? 
51615.801 0.835 -14.01  65.8 -152  ? 
51615.842 0.836 -14.39  65.8 -146  ? 
51616.623 0.857 -14.39  60.0 -163  ? 
51617.608 0.883 -13.96  58.7 -138  ? 
51817.984 0.262 -13.96  35.3  193  ? 
51818.977 0.288 -14.02  33.6  158  77
51819.980 0.315 -14.27  35.5  144  96
51820.996 0.342 -14.20  37.9  140 100
51821.996 0.369 -13.28  40.9  140 117
51822.982 0.396 -14.00  44.4  133 122
51823.971 0.422 -13.46  46.9  131 126
51824.981 0.449 -12.68  53.2  123 126
51830.997 0.611 -14.50  57.6  -96  ? 
51849.861 0.117 -15.69  43.4  175  ? 
51849.914 0.119 -15.47  40.1  179  ? 
51849.961 0.120 -14.25  39.9  177  ? 
51851.046 0.149 -17.06  39.3  181  ? 
51851.055 0.149 -16.71  40.9  183  ? 
51851.839 0.170 -16.05  38.9  184  ? 
51851.854 0.171 -15.72  35.1  184  ? 
51889.975 0.194 -15.07  29.4  187  ? 
51890.901 0.219 -14.71  34.3  203  ? 
51892.904 0.273 -14.60  31.6  195  ? 
51893.902 0.299 -14.78  32.1  161  83
51894.901 0.326 -14.63  35.1  147  98
51895.969 0.355 -14.29  40.9  144 122
51896.942 0.381 -14.66  40.7  131 135
51897.937 0.408 -14.56  46.4  126 141
51898.948 0.435 -14.27  44.7  116 135
51899.942 0.461 -13.67  48.4  127 143
51900.935 0.488 -13.71  53.7  112 160
51901.944 0.515 -13.92  51.0   11  ? 
51913.671 0.830 -13.54  72.2 -160  ? 
51913.717 0.831 -13.25  72.8 -164  ? 
51913.761 0.832 -13.98  72.2 -160  ? 
51913.807 0.834 -14.11  72.6 -162  ? 
51913.831 0.834 -14.14  71.6 -160  ? 
51913.851 0.835 -14.19  71.5 -162  ? 
51913.931 0.837 -13.91  72.0 -160  ? 
51913.969 0.838 -13.81  69.8 -160  ? 
51914.014 0.839 -12.79  69.0 -159  ? 
51914.706 0.858 -14.23  66.8 -158  ? 
51914.808 0.860 -14.34  68.7 -157  ? 
51915.693 0.884 -13.88  66.5 -154  ? 
51915.709 0.885 -13.94  66.8 -152  ? 
51915.766 0.886 -13.82  67.5 -151  ? 
51916.919 0.917 -14.32  64.7 -147  ? 
51916.960 0.918 -14.32  60.8 -147  ? 
51916.976 0.919 -14.22  60.9 -152  ? 
51917.757 0.940 -13.90  57.3 -144  ? 
53291.988 0.825 -15.21  52.0  -73  98
53293.014 0.853 -15.12  49.4  -64 107
53294.994 0.906 -15.09  41.1  -43 132
53294.996 0.906 -15.00  38.1  -45 132
54020.984 0.392 -13.53  31.7   ?   ? 
54024.967 0.499 -13.95  39.9   ?   ? 


\end{verbatim}

\end{document}